\documentclass[onecolumn]{pasj02}
\usepackage[switch,mathlines]{lineno} 
\usepackage{url}
\usepackage{graphicx}
\usepackage{overpic}

\jyear{2026}
\Received{}
\Accepted{}

\begin{document} 

\title{Systematic Study of Cosmic Rays in Star-Forming and Starburst Galaxies using Fermi-LAT and Multiwavelength Data}

\author{
 Hinako \textsc{Ochi},\altaffilmark{1}
 Tsunefumi \textsc{Mizuno},\altaffilmark{2}$^{,\dag}$\orcid{0000-0001-7263-0296}
 \email{mizuno@astro.hiroshima-u.ac.jp} 
 Yasushi \textsc{Fukazawa},\altaffilmark{1}\orcid{0000-0002-0921-8837}
 Noemi \textsc{Barnier},\altaffilmark{3}\orcid{0000-0002-4180-174X} 
 Susumu \textsc{Inoue},\altaffilmark{3}\orcid{0000-0003-1096-9424} 
 Wataru \textsc{Iwakiri},\altaffilmark{3}\orcid{0000-0002-0207-9010} 
 Junling \textsc{Chen},\altaffilmark{4}\orcid{0009-0008-0756-2117} 
 and 
 Tomonori \textsc{Totani}\altaffilmark{4,5}\orcid{0000-0002-6156-6783} 
}
\altaffiltext{1}{Department of Physics, Graduate School of Advanced Science and Engineering, Hiroshima University, 1-3-1 Kagamiyama, Higashi-Hiroshima, Hiroshima 739-8526, Japan}
\altaffiltext{2}{Hiroshima Astrophysical Science Center, Hiroshima University, 1-3-1 Kagamiyama, Higashi-Hiroshima, Hiroshima 739-8526, Japan}
\altaffiltext{3}{International Center for Hadron Astrophysics, Chiba University, 1-33 Yayoi-cho, Inage-ku, Chiba, Chiba 263-8522, Japan}
\altaffiltext{4}{Department of Astronomy, School of Science, the University of Tokyo, Bunkyo-ku, Tokyo 113-0033, Japan}
\altaffiltext{5}{Research Center for the Early Universe, School of Science, the University of Tokyo, Bunkyo-ku, Tokyo 113-0033, Japan}




\maketitle

\begin{abstract}
The majority of $\gamma$ rays from star-forming galaxies (SFGs) and starburst galaxies (SBGs) are expected to originate from interactions between high-energy cosmic rays (CRs) and the interstellar medium, making $\gamma$-ray observations a powerful tool for probing CR populations and interstellar environments. Using 15 years of Fermi-LAT data, we conducted a systematic study of seven sources. The $\gamma$-ray spectra of the bright SBGs M82 and NGC~253 exhibit a decrease in energy flux below a few hundred MeV, suggesting a substantially smaller contribution from CR electrons than previously inferred. Detailed fits of the spectral energy distribution, which combine our GeV $\gamma$-ray measurements with reported radio and TeV $\gamma$-ray results, indicate that the electron-to-proton ratios in M82 and NGC~253 are comparable to that of the Milky Way. 
In contrast, energy densities of CRs in these two SBGs are 
2.5--8 times lower than that of the magnetic field, unlike the Milky Way.
For the nearby Local Group galaxy M31, we decomposed the $\gamma$-ray emission into two components, revealing a central source with a soft spectrum and a compact size of less than $0\fdg33$, which indicates a considerable contribution from unresolved sources.
Our new analysis also reduces the $\gamma$-ray fluxes of several SBGs, bringing all SBG sample members to levels comparable to or below the calorimetric limit, with calorimetric efficiencies ranging from 0.33 to 1.20. In contrast, all SFG sample members exhibit low efficiencies 
at the 10\% level. 
We developed a theoretical model of interstellar $\gamma$-ray emission 
that successfully reproduces the observed
calorimetric efficiencies,
providing a robust framework for interpreting and predicting $\gamma$-ray emission from SFGs and SBGs.
\end{abstract}


\section{Introduction}
Cosmic rays (CRs) in the Milky Way Galaxy are believed to originate from supernova remnants and other star-formation processes, such as the winds of massive stars and pulsars
(e.g., \cite{Ginzburg1964,Butt2009}). Similarly, galaxies with high star-formation rates are expected to exhibit CR densities comparable to or greater than those of the Milky Way.
In these external galaxies, CRs generate electromagnetic radiation through mechanisms analogous to those in the Milky Way. 
Specifically, CR electrons produce synchrotron radiation via interactions
with interstellar magnetic fields and non-thermal Bremsstrahlung radiation via interactions with interstellar gas. They also upscatter the interstellar radiation field (ISRF) via
inverse Compton (IC) scattering. CR protons produce $\gamma$ rays through nuclear--nuclear collisions with interstellar gas. 
Multiwavelength observations of external galaxies therefore provides insights into both CR populations and the properties of their interstellar environments,
including magnetic fields, the ISRF, and gas content, thereby advancing understanding of the underlying astrophysical processes.

The $\gamma$-ray band is particularly important because it is sensitive to CR protons within galaxies. CR protons produce intense GeV and TeV $\gamma$ rays through the decay of neutral pions, which are generated by nuclear--nuclear interactions with ambient interstellar gas. This process typically results in $\gamma$-ray emission that dominates over that from CR electrons in 
typical interstellar environments.
Fermi-LAT \citep{Atwood2009} detected GeV $\gamma$-rays from the nearest star-forming galaxies (SFGs), specifically the LMC, SMC, and M31 \citep{FermiLMC,FermiSMC,FermiM31}, shortly after its launch.
The instrument also identified significant GeV $\gamma$-ray emission from the nearby starburst galaxies (SBGs) M82 and NGC~253 \citep{FermiM82N253}.
Concurrently, ground-based Cherenkov telescopes observed significant TeV $\gamma$-rays from these two galaxies \citep{VERITAS_M82,HESS_N253}, thereby establishing 
SBGs as $\gamma$-ray sources. A strong correlation between observed GeV $\gamma$-ray luminosities and total infrared luminosities \citep{FermiSFG} suggests that the 
majority of $\gamma$ rays originate from interstellar CRs produced by star-formation activity, rather than from the nuclear supermassive black hole. 
The hard spectrum observed in SBGs (photon index $\Gamma \sim 2.2$) indicates that these galaxies act as CR calorimeters, in contrast to the Milky Way, which is escape-dominated \citep{FermiPaper2}. 
These observations have prompted theoretical investigations into CR transport in both SBGs and the Milky Way. For example, \citet{Krumholz2020} proposed that CR diffusion in SBGs is nearly 
energy-independent up to about 1~TeV. Furthermore, the absence of a decrease in $\gamma$-ray flux below a few hundred MeV in M82, which would be expected in a proton-dominated scenario, suggests a high
CR electron-to-proton ratio ($K_\mathrm{ep}$), as indicated by VERITAS 2025 \citep{VERITAS_M82v2}.

Understanding $\gamma$ rays produced by star-formation activity is also essential for investigating the activity of an active galactic nucleus (AGN).
When galactic nucleus emission does not dominate the observed $\gamma$ rays, as is typical for non-jetted AGNs,
the $\gamma$-ray contributions from the nucleus and from star-formation activity are expected to be comparable.
Current and near-future GeV/TeV $\gamma$-ray instruments are not able to resolve nuclear emission from large-scale emission except in local group galaxies.
However, differences in spectral shape can enable their distinction.
For example, \citet{Eichmann2022} presented a detailed model of multi-wavelength data for the Seyfert galaxy NGC~1068, noting that coronal emission from the Seyfert nucleus would appear below 100~MeV. Indeed,
\citet{Ajello2023} reported significant $\gamma$-ray emission below a few hundred MeV from 
the galaxy using the Fermi data.
The observed flux exceeds the expected starburst component in a proton-dominated scenario, leading the authors to attribute the emission to MeV radiation from the corona surrounding the Seyfert nucleus.

In short, accurate measurements of $\gamma$ rays from SFGs and SBGs are essential for understanding CRs and the interstellar environments
of these systems. Furthermore, improved knowledge of SFGs and SBGs enables
robust identification and characterization of coronal emission in nearby Seyfert galaxies --- many of which exhibit starburst activity --- relative to the large-scale starburst components. Recent improvements in the Fermi-LAT instrument response 
(Pass~8; \cite{P8ref1,P8ref2}),
together with more than 15 years of accumulated observations, 
now make it possible to study $\gamma$-ray emission from SFGs, SBGs, and Seyferts with unprecedented accuracy and uniformity.
Motivated by these scientific and technical advances, we have carried out a systematic analysis of individually detected SFGs and SBGs with the Fermi-LAT.
The methodology and results are detailed in this paper.
A dedicated analysis of Seyfert galaxies,
including a comprehensive comparison with SFGs and SBGs, will be reported separately (Barnier et al. 2026, in preparation).
This paper is organized as follows.
Section~2 describes the target selection, $\gamma$-ray observations, and analysis procedures.
Section~3 presents the results for each galaxy.
Section~4 discusses the CR properties inferred from detailed modeling of the multi-wavelength spectra from radio to TeV $\gamma$ rays, 
as well as comparisons with theoretical calculations of the GeV $\gamma$ rays and among our targets.
Section~5 summarizes the findings and outlines future prospects.

\clearpage
\section{Target Selection and Observations}

\subsection{Target Selection}
We aim to accurately characterize the $\gamma$-ray spectra of SFGs and SBGs and therefore focus on sources that are individually detected by Fermi-LAT.
For source selection, we used the 12-year Fermi-LAT source catalog (4FGL-DR3\footnote{
\url{https://fermi.gsfc.nasa.gov/ssc/data/access/lat/12yr_catalog/}
};\cite{4FGLDR3}),
which contains more than 6,000 $\gamma$-ray sources. 
The catalog provides source classifications: six objects are listed as normal galaxies,
eight as starburst galaxies, and four as Seyfert galaxies. 
In general, galaxy types can be distinguished using their broadband spectral properties from radio to X-ray energies. 
Systems with high radio to optical fluxes are typically jetted AGN (radio galaxies, blazars, etc.). 
Among galaxies with relatively weak radio emission, those with strong infrared emission are classified as SFGs or SBGs,
while those with strong hard X-ray emission are identified as Seyfert galaxies. 
From the 14 objects labeled as 
``normal galaxies'' or ``starburst galaxies'' in the 4FGL, 
we selected six as our SFG/SBG samples and analyzed them in this work. The other eight galaxies are excluded for the following reasons:
\begin{itemize}
\item Four relatively faint systems with photon flux in 1~GeV to 100~GeV $f_\mathrm{ph, \mbox{\scriptsize{1--100}}} \le 10^{-10}~\mathrm{photons~s^{-1}~cm^{-2}}$:
we excluded them because they are not suitable for a detailed analysis of $\gamma$-ray emission.
\item NGC~1068 and NGC~4945: although they
are labeled as starbursts in 4FGL due to their high infrared luminosity (reflecting high star-formation rates),
both exhibit strong hard X-ray emission indicative of significant nuclear activity, and have been classified as Seyfert galaxies in several previous Fermi-LAT
studies (e.g., \cite{Ajello2023}). We follow this convention
and defer their analysis to the Seyfert-focused companion paper (Barnier et al. 2026, in preparation).
\item The LMC and SMC: they are nearby $\gamma$-ray bright galaxies, but their large angular extent
complicates the analysis. For simplicity, we excluded them from our target list for the analysis,
and instead use their published 
$\gamma$-ray fluxes in the discussion (Section~4.3).
\end{itemize}
Instead, we additionally included M33, previously studied as an SFG by \citet{Ajello2020}.
Our final target list is summarized in Table~1.
All our sources were already analyzed in \citet{Ajello2020}, 
but our dedicated spectral and spatial modeling (for M82, NGC~253, and M31; see Sections~3.1--3.3)
and use of more data (10 yr vs. 15 yr) and newer Galactic diffuse and point-source models
enable improved characterization of the $\gamma$-ray emission.
We adopted the M Ajello et al. (2020) distances
\footnote{
In Section~4, we evaluate interstellar environmental parameters based on the literature to model the multi-wavelength spectrum. For consistency, we scale the distance-dependent quantities (such as luminosities) to match the distance we adopted.
}
for easy comparison.

\begin{table}[htbp]
  \tbl{List of SFGs/SBGs analyzed in this paper}{%
  \begin{tabular}{cccc}
    \hline
Common name & 4FGL name & Distance\footnotemark[$*$] & (RA, DEC)$_{\rm J2000.0}$\footnotemark[$\dagger$] \\ 
            &           & Mpc & \\
\hline
M82      & 4FGL~J0955.7+6940 &  3.7  & (148\fdg95, +69\fdg67) \\
NGC~253  & 4FGL~J0047.5-2517 &  3.3  & ( 11\fdg90, -25\fdg29) \\
M31      & 4FGL~J0043.2+4114 &  0.78 & ( 10\fdg82, +41\fdg24) \\
Arp~220  & 4FGL~J1534.7+2331 & 79.9  & (233\fdg70, +23\fdg53) \\
NGC~2146 & 4FGL~J0618.1+7819 & 18.0  & ( 94\fdg53, +78\fdg33) \\
Arp~299  & 4FGL~J1128.2+5831 & 47.7  & (172\fdg14, +58\fdg56) \\
M33      & --                &  0.85 & ( 23\fdg61, +30\fdg78) \\
    \hline
  \end{tabular}}\label{tab:first}
\begin{tabnote}
\footnotemark[$*$] 
Distances are taken from \citet{Ajello2020}. \\ 
\footnotemark[$\dagger$] 
RA and DEC are right ascension and declination, respectively, and taken from 4FGL. See
\S~3.4 for M33. Also see \S~3.3 for modeling the spatial distribution of $\gamma$ rays from M31.\\ 

\end{tabnote}
\end{table}

\clearpage
\subsection{$\gamma$-ray Observation and Analysis Procedure}
Fermi-LAT \citep{Atwood2009} is a pair-tracking $\gamma$-ray telescope sensitive to photons 
from {$\sim$}20~MeV to beyond 300~GeV.
Its wide field of view (${\sim}$2.4~sr) and continuous sky-survey operation
--- covering nearly the entire sky every three hours ---
make it an ideal instrument for systematic studies of SFGs/SBGs 
(as well as other source classes) in the sub-GeV to GeV band.
Routine LAT science operations began on August 4, 2008.
For this work, we analyzed 15 years of data collected from
August 4, 2008 to August 2, 2023.
The analysis was performed using the standard LAT analysis software package Fermitools\footnote{
\url{https://fermi.gsfc.nasa.gov/ssc/data/analysis/software/}
} v2.2.0,
together with Fermipy\footnote{
\url{https://fermipy.readthedocs.io/en/latest/}
} v1.3.1, a Python-based pipeline that automates many LAT analysis tasks
\citep{Wood2017}.

We defined a $15\arcdeg \times 15\arcdeg$ region of interest (ROI) 
and selected events belonging to the SOURCE class 
to maximize the photon statistics.
We set the ROI aligned in Galactic coordinates to reduce potential
contamination from Galactic diffuse emission, but will show the resulting maps
in equatorial coordinates for easy comparison with past studies.
Data from 50~MeV to 102.4~GeV were binned spatially with a pixel size of $0\fdg1$.
As described below, we considered logarithmically spaced energy intervals, each increasing by a factor of two (e.g., 50--100~MeV, 100-200~MeV, and so forth
until 5120--10240~MeV) in the analysis.

In order to reduce contamination from $\gamma$ rays produced in Earth's atmosphere,
we applied zenith-angle cuts of $80^{\circ}$, $90^{\circ}$, and $105^{\circ}$ for events in the 50--100~MeV, 100--400~MeV, and {$\ge$}400~MeV bands, respectively.
These thresholds account for the LAT's relatively poor angular resolution at low energies (68\% containment radius, $r_{68}$, is ${\sim}5^{\circ}$ at 100~MeV).
Similarly, we restricted the event selection to PSF types 2 and 3 (the second-best and best angular resolution) for 50--100~MeV,
and those of PSF types 1, 2, and 3 for 100--400~MeV. Above 400~MeV, where the angular resolution is significantly better
($r_{68}$ $\le 2^{\circ}$),
we included all PSF event types.

For spectral analysis, 
we divided each energy interval into three bins
(i.e., about 10 bins per decade)
to capture the effect of finite energy dispersion.
We computed the binned maximum likelihood for our ROI using the
P8R3\_SOURCE\_V3 instrument response functions, which match our event selections. 
Below 400~MeV, we employed the ``summed likelihood'' method in Fermitools,
treating each PSF type independently to best use the angular resolution and 
better separate the target galaxy from nearby sources. 
Above 400~MeV, where confusion is less severe,
we combined all PSF types for simplicity.
Our model includes the standard Galactic diffuse emission model (gll\_iem\_v07.fits), the isotropic spectral template (iso\_P8R3\_SOURCE\_V3\_v1.txt), 
and all point-source models from the 4FGL-DR3 catalog \citep{4FGLDR3}.
For point sources within $7\fdg5$ of the ROI center,
we allowed both the normalization and spectral index to vary freely.
We also left free the normalization and spectral index of the 
Galactic diffuse background model, as well as the normalization 
of the isotropic spectral template.
The Sun and Moon are known $\gamma$-ray emitters whose motion across the
sky can affect LAT analyses if they pass near the ROI.
For all targets in our sample, the ecliptic plane lies outside the ROI,
and the solar and lunar emissions in the ROI are both weak and very extended in these configurations.
Therefore their contribution will be absorbed by the isotropic component,
and we did not include templates for the Sun or Moon.

The spectrum of each galaxy below a few hundred MeV is essential for constraining  $K_\mathrm{ep}$, where the LAT's angular resolution 
is relatively poor.
An inadequate global spectral model can distort the modeling of nearby sources
and bias the derived spectral energy distribution (SED) of the target source.
To mitigate this, we adopted two alternative spectral models
--- the power-law (PL) and log-parabola (logP) ---
for bright systems (M82, NGC~253, and M31).

The differential photon flux ($dN/dE$) as a function of energy ($E$) of the
PL model is given by
\begin{equation}
\frac{dN}{dE} = K \left( \frac{E}{E_{0}} \right)^{-\Gamma}~,
\end{equation}
where $K$ is the normalization, $E_{0}$ is the reference energy, and $\Gamma$ is the photon index. The logP model is expressed as
\begin{equation}
\frac{dN}{dE} = K \left( \frac{E}{E_\mathrm{b}} \right)^{-(\alpha+\beta \log(E/E_\mathrm{b}))}~~,
\end{equation}
where $E_\mathrm{b}$ is the reference energy, and $\alpha$ end $\beta$ describe the spectral 
slope and curvature, respectively.
The logP model is widely used to represent curved $\gamma$-ray spectra.
When applying the logP model,
we fixed $E_\mathrm{b}$ to the value listed in the 4FGL catalog.
If the 4FGL catalog uses the PL model,
we adopted $E_\mathrm{b} = 1000~\mathrm{MeV}$, typical of a Milky-Way-like emission.
After obtaining the best-fit global spectral model, we derived
the SED by fixing all parameters except the normalization of the target source
and refitting the normalization independently in each energy interval (50--100~MeV, 100--200~MeV, etc.).
We then compared the resulting SED with the best-fit global spectral model
and selected the model function (PL or logP) that provides the better overall agreement.

To evaluate the detection significance in each energy interval, we
used the test statistic (TS), defined as $\mathrm{TS}=2\left( \ln{L_{1}}-\ln{L_{0}} \right)$,
where $L_{1}$ and $L_{0}$ are the maximum likelihood values with and without the source, respectively.
The TS approximately follows a $\chi^{2}$ distribution with degrees of freedom
equal to the number of free parameters \citep{Mattox1996}. 
For intarvals in which $\mathrm{TS<4}$ (corresponding to ${\le}2\sigma$ significance),
we report 95\% upper limits.

\clearpage
\section{Data Analysis of Each Object}

\subsection{M82}

Figure~1 (panels (a) and (c)) presents the count maps of M82, constructed separately for energies above and below 400~MeV.
The source is distinctly detected in the high-energy band and remains identifiable in the low-energy band.
As outlined in Section~2, two spectral models were employed:
the logP model (as adopted in 4FGL) and the PL model.
The PL model yields 
$\Gamma= 2.145 \pm 0.027$, while the logP model provides $\alpha = 2.021 \pm 0.046$ and 
$\beta = 0.110 \pm 0.024$, with $E_\mathrm{b}$ fixed at 956~MeV. 
Our best-fit PL index is almost identical to that of \citet{Ajello2020} ($\Gamma=2.14$),
who employed only the PL model because the curvature was not significant due to low statistics.
Our logP model parameters are similar to those of 4FGL
($\alpha = 2.115$ and $\beta = 0.074$) and require
a moderately curved spectrum.
The SEDs were evaluated as described in Section~2 and compared with the best-fit global models in Figure~2. The best-fit PL model does not adequately reproduce the SED, whereas the logP model shows good agreement.
Comparison of SEDs based on the PL and logP models
indicates that use of the PL model artificially increases the low-energy flux
of M82,
particularly below a few hundred MeV,
by reducing the low-energy fluxes in models for diffuse emission and/or nearby sources.

These results indicate that the GeV and sub-GeV emission from M82 exhibits
mild curvature\footnote{
Although \citet{Persic2025} reported a hint of a rising spectrum below a few hundred MeV using all PSF event types, we continue to use tighter cuts to reduce possible contamination from nearby sources or diffuse emission. We note that they obtained a spectrum similar to ours with a more stringent cut.
}
, and use of the PL model introduces bias in the low-energy flux. 
This bias will affect the SED fitting described in Section~4.1.1. 
Therefore, the logP model was adopted for subsequent analysis (this Section) and discussion (Section~4).
The source is detected 
at a TS of 1702, corresponding to a significance greater than $40\sigma$. 
The resulting model count maps are shown in Figure~1 (panels (b) and (d)), 
and the spectral parameters are summarized in Table~2. 
Numerical values of the SED are provided in Table~3.

\vspace{0.5cm}
\begin{figure}[htbp]
\begin{tabular}{cc}
\begin{minipage}{\textwidth}
\centering
\begin{overpic}[width=\textwidth]{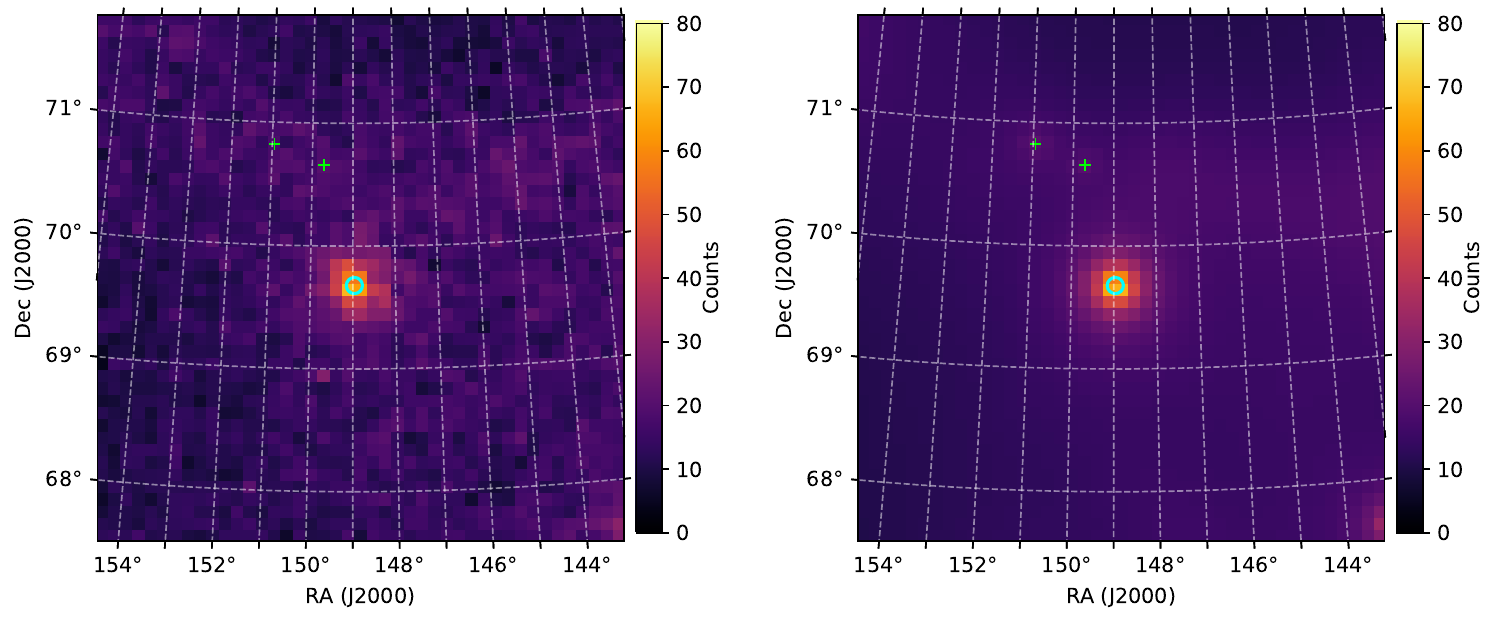}
\put(7,42){(a) M82 ($\ge$400~MeV)}
\put(57,42){(b) M82 ($\ge$400~MeV)}
\end{overpic}
\end{minipage}
\\
\\
\\
\begin{minipage}{\textwidth}
\centering
\begin{overpic}[width=\textwidth]{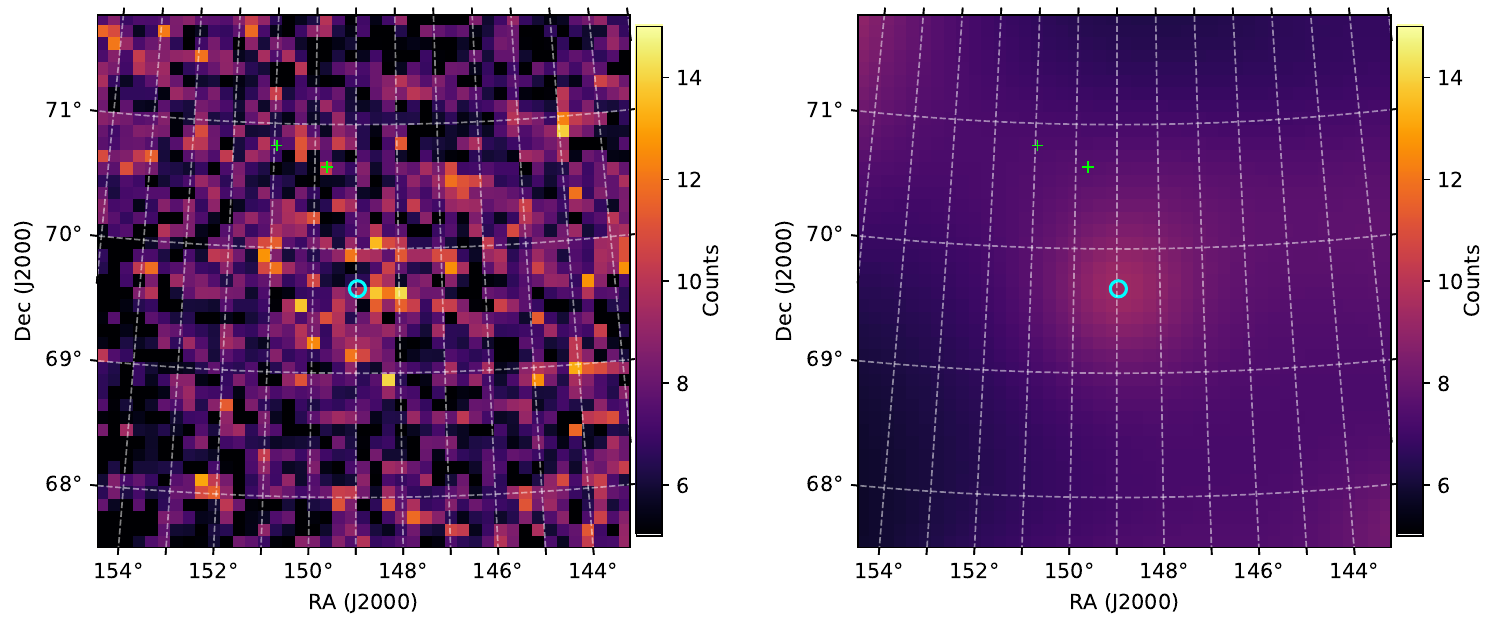}
\put(7,42){(c) M82 ($\le$400~MeV)}
\put(57,42){(d) M82 ($\le$400~MeV)}
\end{overpic}
\end{minipage} 
\vspace{0.5cm}
\end{tabular}
\caption{(a) Data count map and (b) model count map in the vicinity of M82 in {$\ge$}400~MeV.
(c) Data count map and (d) model count map in a 
$4\arcdeg \times 4\arcdeg$ field of M82 
in {$\le$}400~MeV.  
In all panels, the position of M82 is indicated by a cyan circle,
and those of 4FGL sources are indicated by green crosses.
{Alt text: Two data count maps and two model count maps for the M82 region.}
}
\end{figure}

\begin{figure}[htbp]
\begin{tabular}{cc}
\begin{minipage}{0.5\textwidth}
\centering
\begin{overpic}[width=\textwidth]{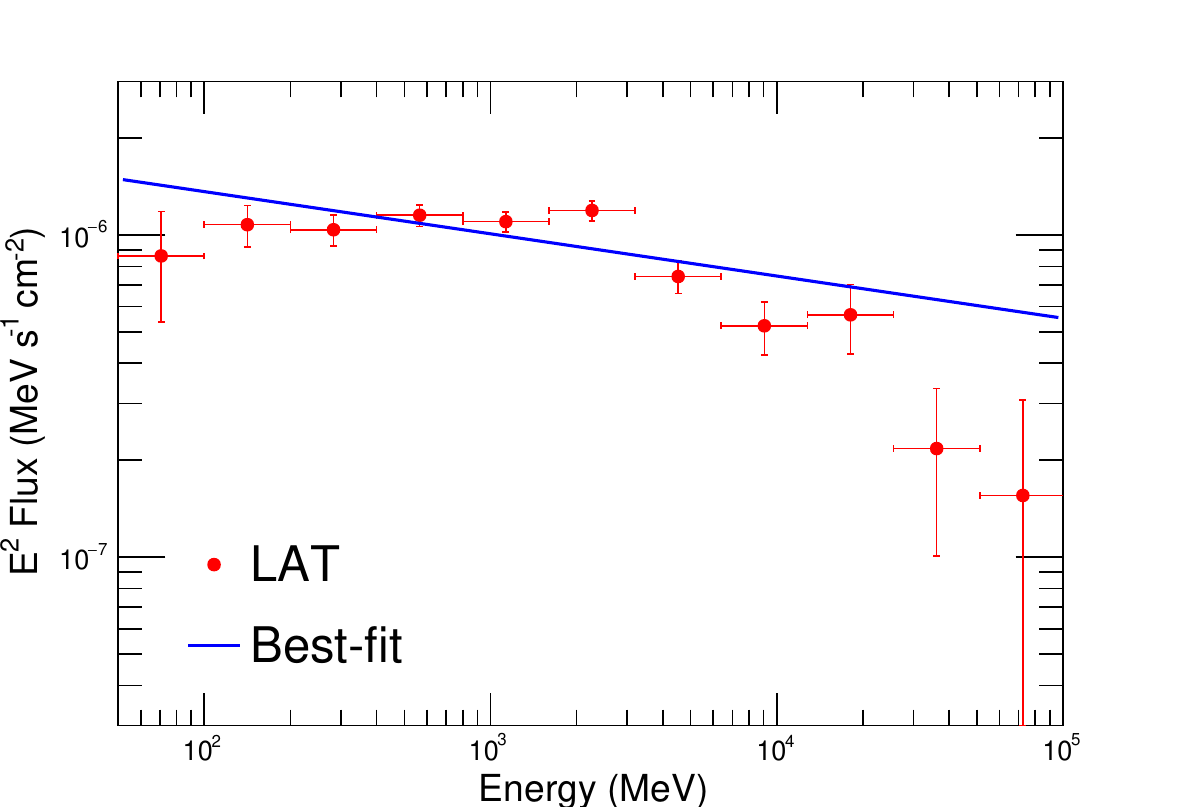}
\put(11,65){(a) M82}
\end{overpic}
\end{minipage}
\begin{minipage}{0.5\textwidth}
\centering
\begin{overpic}[width=\textwidth]{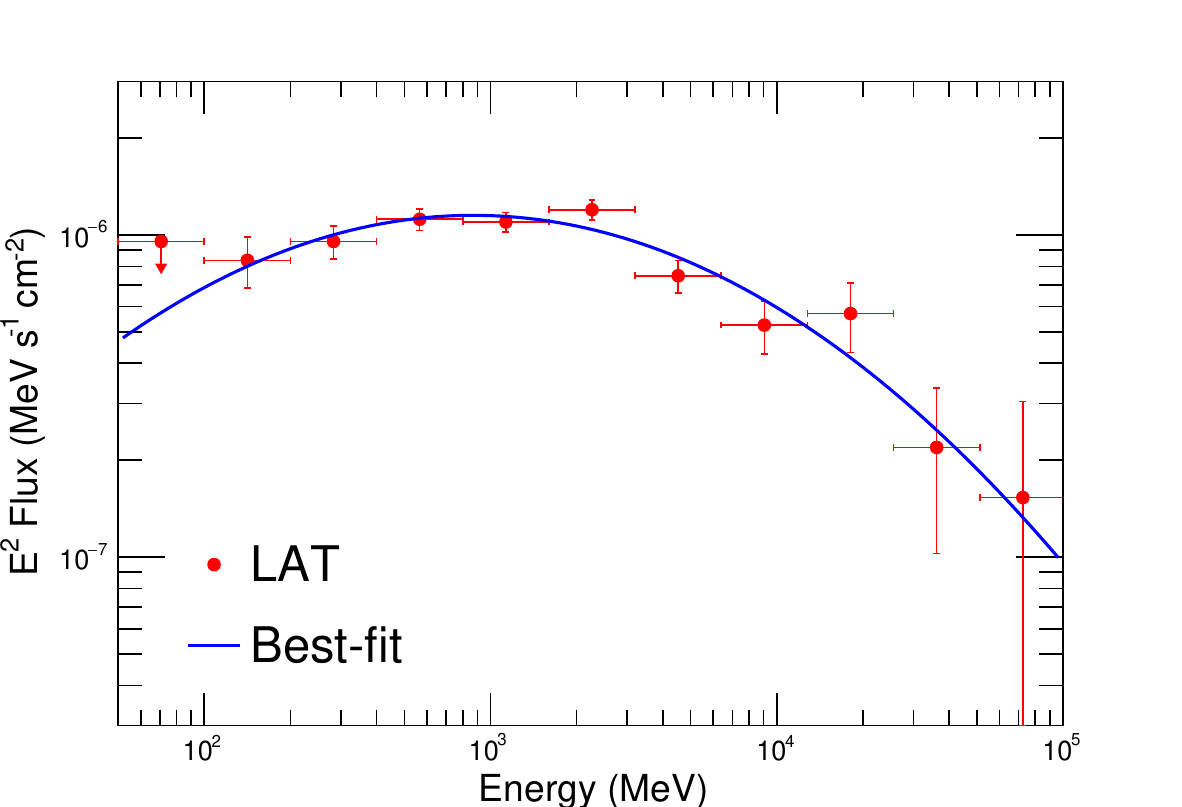}
\put(11,65){(b) M82}
\end{overpic}
\end{minipage} 
\end{tabular}
\\
\caption{SED and a best-fit spectral model of M82, obtained by (a) the PL model and (b) the logP model.
The choice of the spectral model formula affects the SED data points (see text for details), and we adopt
the logP model results in the discussion section.
{Alt text: Two SED plots for M82.}
}
\end{figure}

\clearpage
\subsection{NGC~253}
Figure~3 represents the data and model count maps of NGC~253:
the source is clearly detected above 400~MeV and is identifiable below 400~MeV.
Similar to the analysis of M82, both the logP model (adopted in 4FGL) and the PL model were tested:
the PL model yields 
$\Gamma= 2.105 \pm 0.038$, while the logP model provides $\alpha = 1.991 \pm 0.064$ and 
$\beta = 0.130 \pm 0.035$, with $E_\mathrm{b}$ fixed at 1048~MeV. 
Again, our best-fit PL index is almost identical to that of \citet{Ajello2020} ($\Gamma=2.10$),
and our logP model parameters are similar to those of 4FGL
($\alpha = 2.066$ and $\beta = 0.089$).
As shown by Figure~4, the logP model demonstrated better agreement with the SED data points,
while the use of the PL model artificially increased the low-energy flux.
Like M82, we adopted the logP model for NGC~253 to prevent the bias in low-energy flux
from affecting the SED fitting (Section~4.1.2).
Table~2 summarizes the spectral parameters, including 
TS of 899 (detection significance of about $30\sigma$).
The numerical values of the SED are provided in Table~3.

\vspace{0.5cm}
\begin{figure}[htbp]
\begin{tabular}{cc}
\begin{minipage}{1\textwidth}
\centering
\begin{overpic}[width=\textwidth]{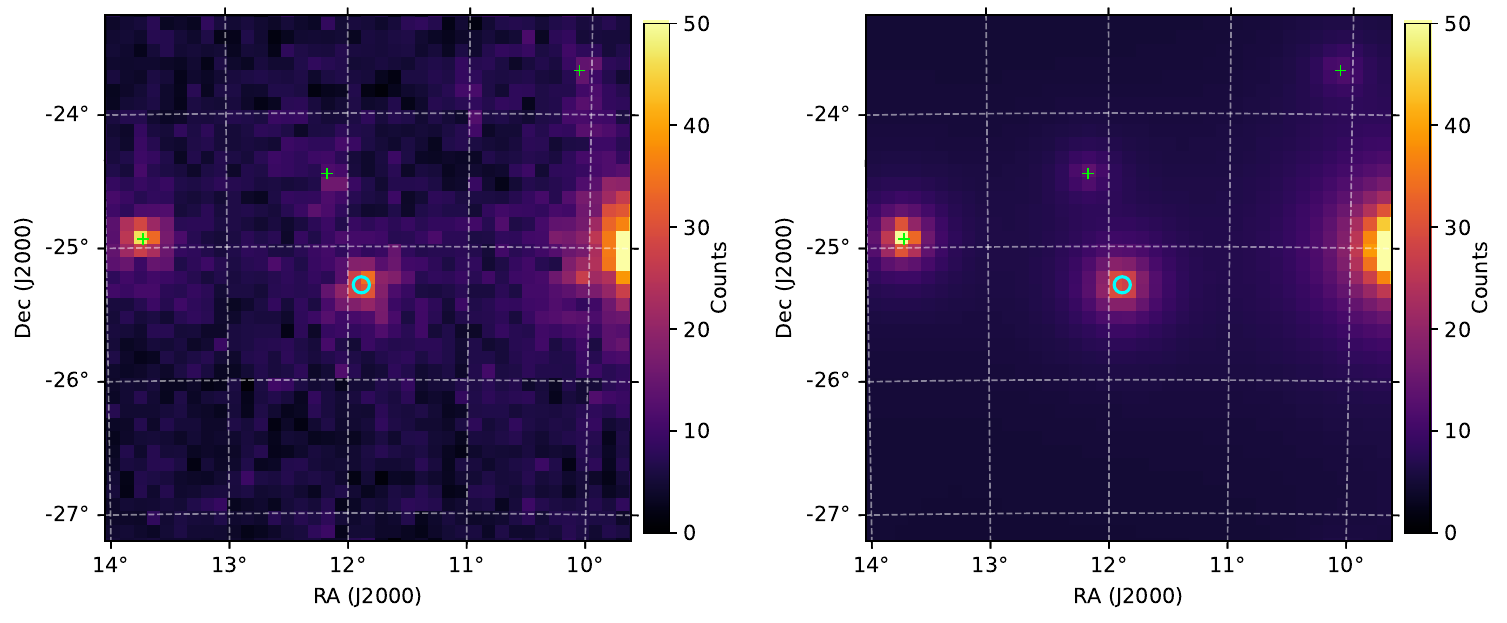}
\put(7,42){(a) NGC~253 ($\ge$400~MeV)}
\put(57,42){(b) NGC~253 ($\ge$400~MeV)}
\end{overpic}
\end{minipage}
\\
\\
\\
\begin{minipage}{1\textwidth}
\centering
\begin{overpic}[width=\textwidth]{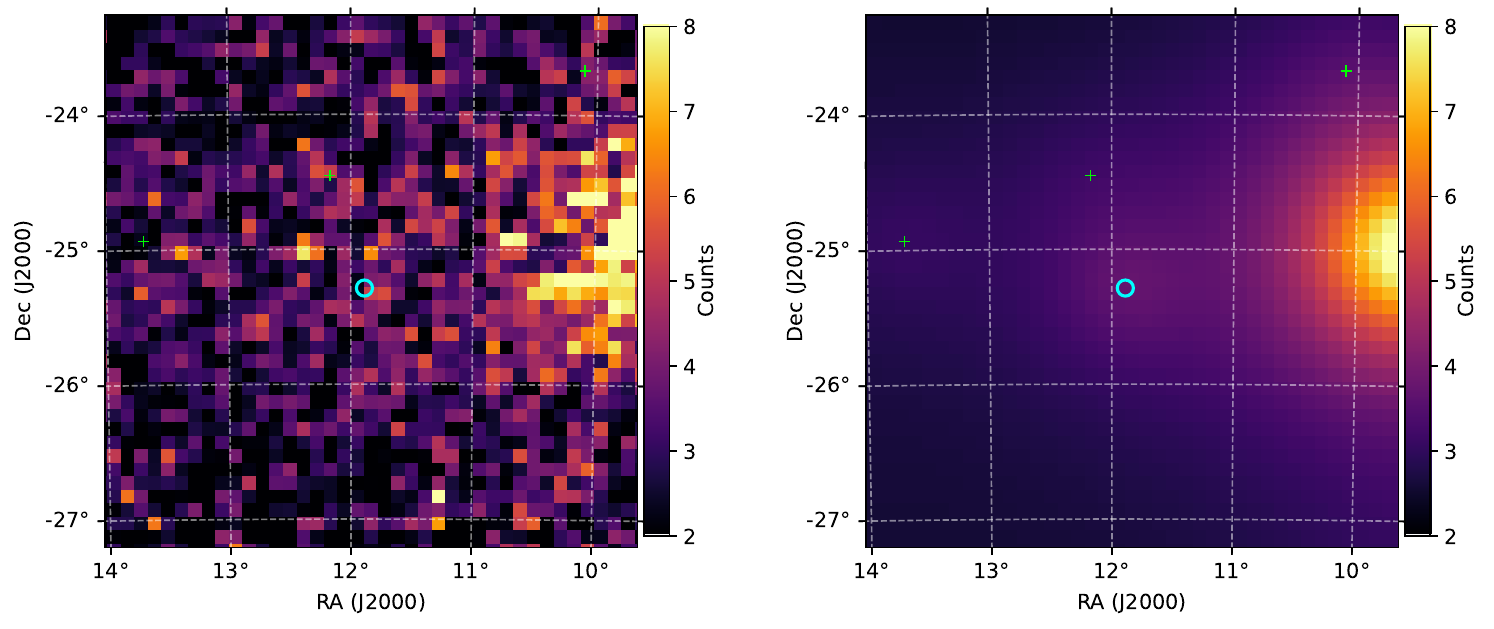}
\put(7,42){(c) NGC~253 ($\le$400~MeV)}
\put(57,42){(d) NGC~253 ($\le$400~MeV)}
\end{overpic}
\end{minipage} 
\vspace{0.5cm}
\end{tabular}
\caption{
(a) Data count map and (b) model count map 
in a 
$4\arcdeg \times 4\arcdeg$ field
of NGC~253 in {$\ge$}400~MeV. (c) Data count map and (d) model count map in the vicinity of NGC~253 in {$\le$}400~MeV.
In all panels, the position of NGC~253 is indicated by a cyan circle, and those of 4FGL sources are indicated by green crosses.
{Alt text: Two data count maps and two model count maps for the NGC~253 region.}
}
\end{figure}

\begin{figure}[htbp]
\begin{tabular}{cc}
\begin{minipage}{0.5\textwidth}
\centering
\begin{overpic}[width=\textwidth]{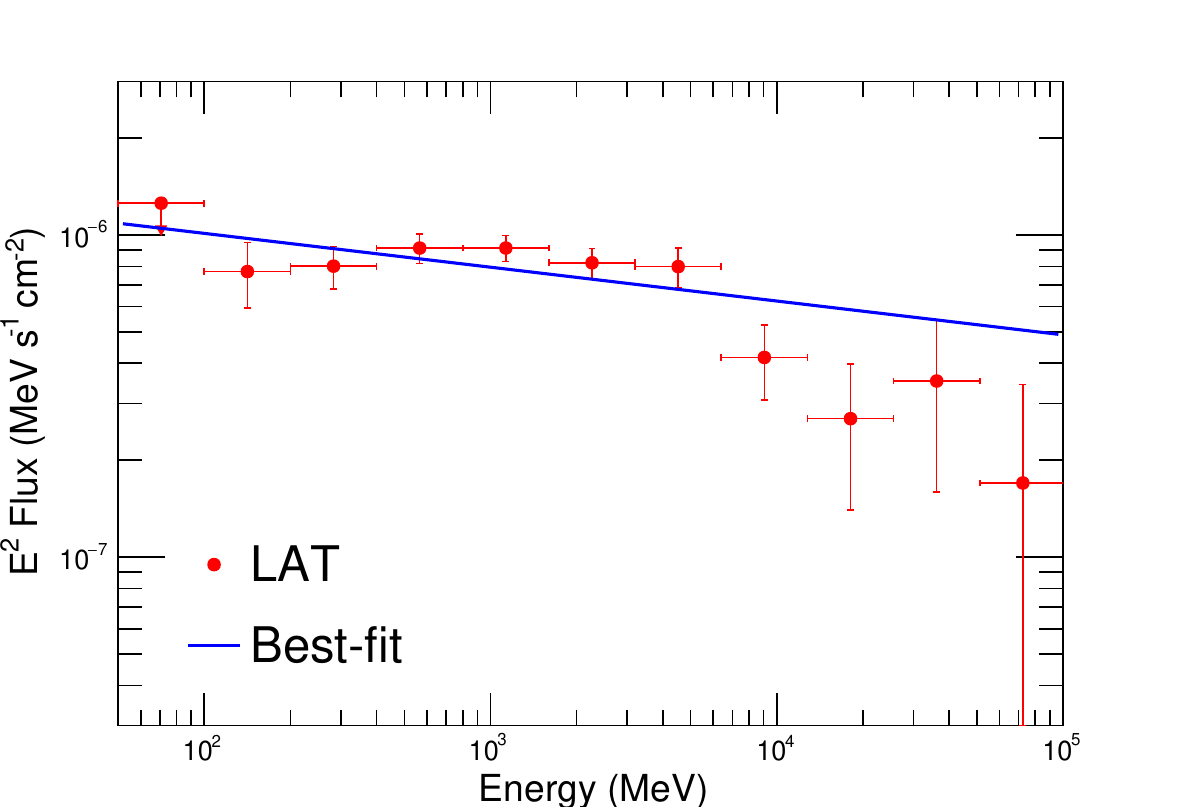}
\put(11,65){(a) NGC~253}
\end{overpic}
\end{minipage}
\begin{minipage}{0.5\textwidth}
\centering
\begin{overpic}[width=\textwidth]{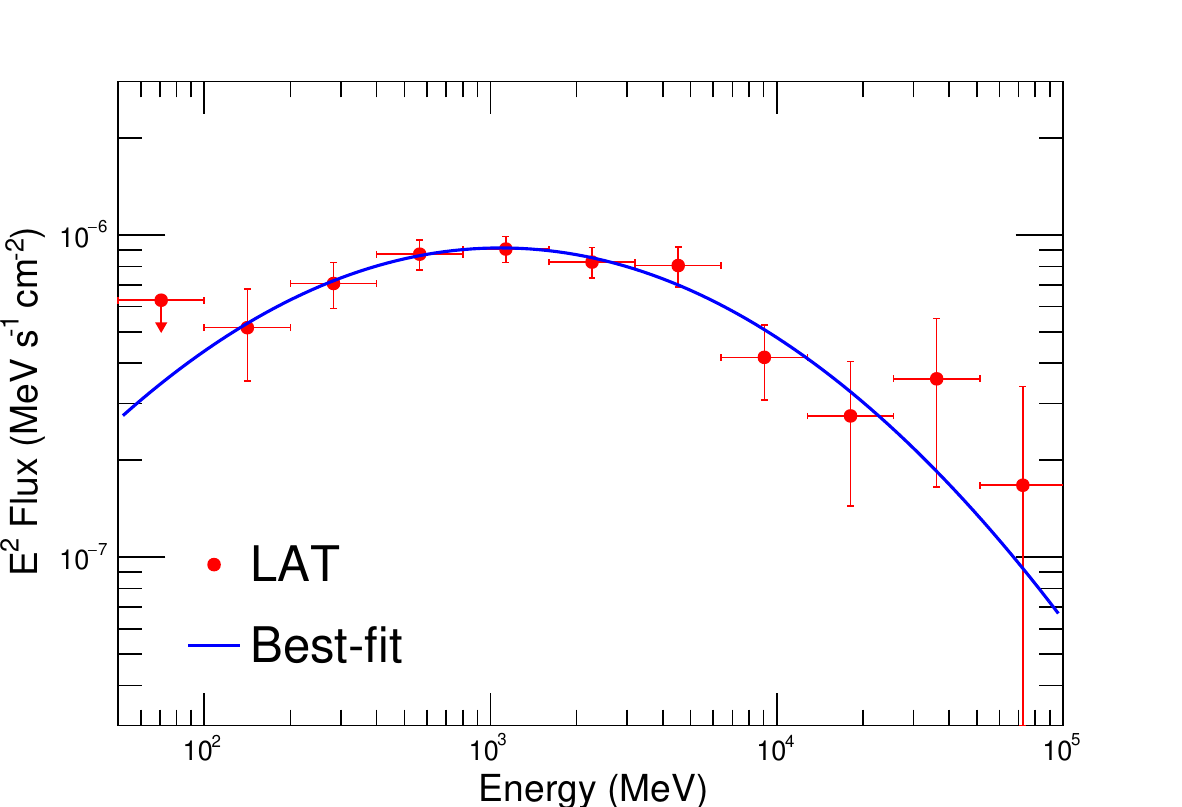}
\put(11,65){(b) NGC~253}
\end{overpic}
\end{minipage} 
\end{tabular}
\caption{
SED and a best-fit spectral model of NGC~253, obtained by (a) the PL model and (b) the logP model.
{Alt text: Two SED plots for NGC~253.}
}
\end{figure}

\clearpage
\subsection{M31}
M31 is a Local Group galaxy located at a distance of {$\sim$0.78~kpc} and therefore spatially resolved even with the moderate angular resolution of the Fermi-LAT (68\% containment radii of
{$\sim$}0\fdg9 and 0\fdg2 at 1~GeV and 10~GeV, respectively).
Using seven years of Fermi-LAT data, \citet{FermiM31v2} reported that the GeV $\gamma$-ray emission is dominated by an extended nuclear component with a radius of ${\sim}0\fdg4$. More recently, 
\citet{Xing2023} showed that the emission is decomposed into two point-like sources: one at the M31 center and another located ${\sim}0\fdg4$ to the southeast (SE).
We confirmed this result by examining TS maps (constructed with the 4FGL M31 source removed) above 1~GeV and 3~GeV (Figure~5).
The two sources are clearly separated in the {$\ge$}3~GeV map, with the central source exhibiting a softer spectrum. 

Accordingly, we modeled M31 emission using two point sources at the positions reported in \citet{Xing2023}. The PL model adequately describes the SE source spectrum
\footnote{If we use the logP model, the best-fit curvature parameter $\beta$ is only 0.03.}
, whereas the central source shows clear spectral curvature and is better represented by the logP model,
as shown in Figure~6.
The comparison between the central and SE source spectra, obtained using the logP and PL models, respectively, is shown in Figure~7.
The spectral parameters and the numerical values of the SED data points are listed in Tables~2 and 3.
The curvature parameter ($\beta$) of the central source is about three times larger than
those of M82 and NGC~253.
The central and SE sources are detected with TS values of {$\sim$}75 ($>8\sigma$) and {$\sim$}44 ($>6\sigma$), respectively. The data and model count maps, generated using the final model configuration, are presented in Figure~8.

We also tested spatial extension of the central source by modeling it as a uniform disk and scanning the radius in steps of $0\fdg01$. To take advantage of the improved angular resolution, we used data above 800~MeV.
The profile of the TS value indicates that the source is consistent with being point-like, and we derived a
95\% upper limit on the radius of $0\fdg33$.

\vspace{0.5cm}
\begin{figure}[htbp]
\begin{tabular}{cc}
\begin{minipage}{1.0\textwidth}
\centering
\begin{overpic}[width=\textwidth]{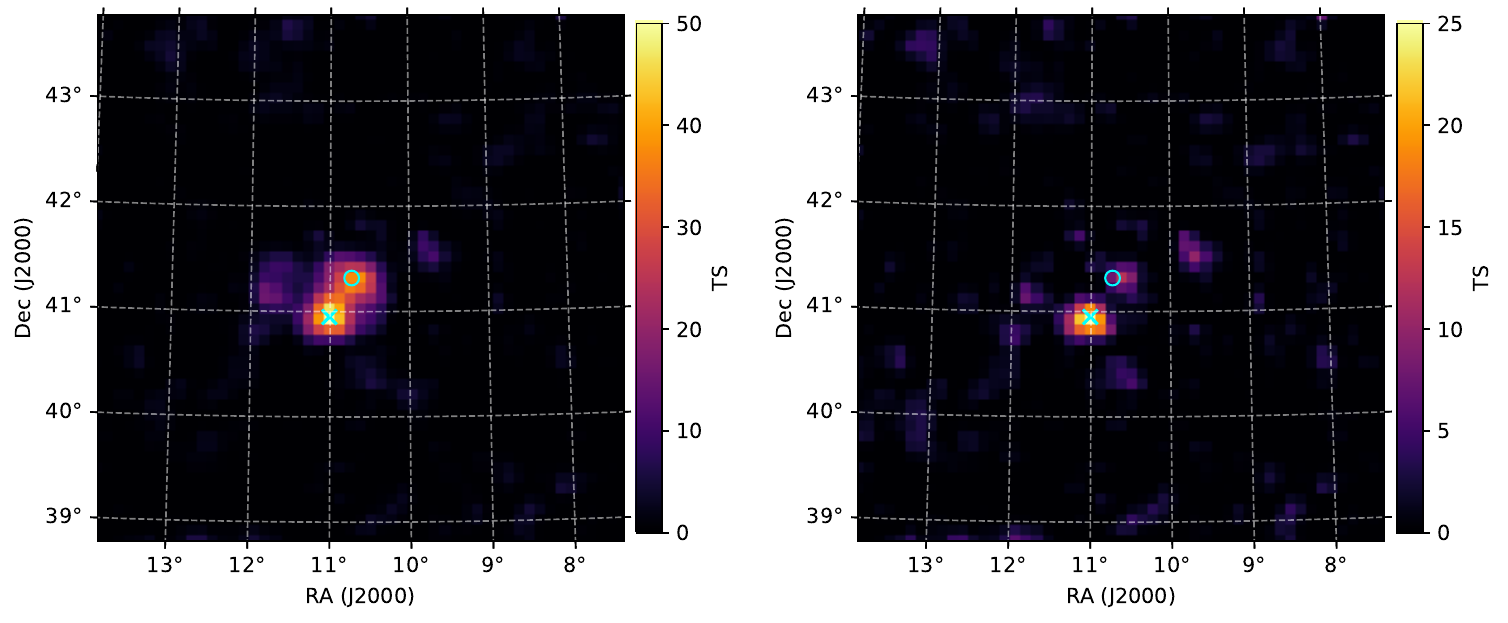}
\put(7,42){(a) M31 ({$\ge$1~GeV})}
\put(57,42){(b) M31 ({$\ge$}3~GeV)}
\end{overpic}
\end{minipage}
\vspace{0.5cm}
\end{tabular}
\caption{TS maps in the vicinity of M31, (a) above 1~GeV and (b) above 3~GeV. Both maps are constructed with the 4FGL source for M31 removed. 
In both panels, the central M31 source is indicated by a cyan circle,
and the SE source is indicated by a cyan cross.
{Alt text: Two TS maps for M31.}
}
\end{figure}

\begin{figure}[htbp]
\begin{tabular}{cc}
\begin{minipage}{0.5\textwidth}
\centering
\begin{overpic}[width=\textwidth]{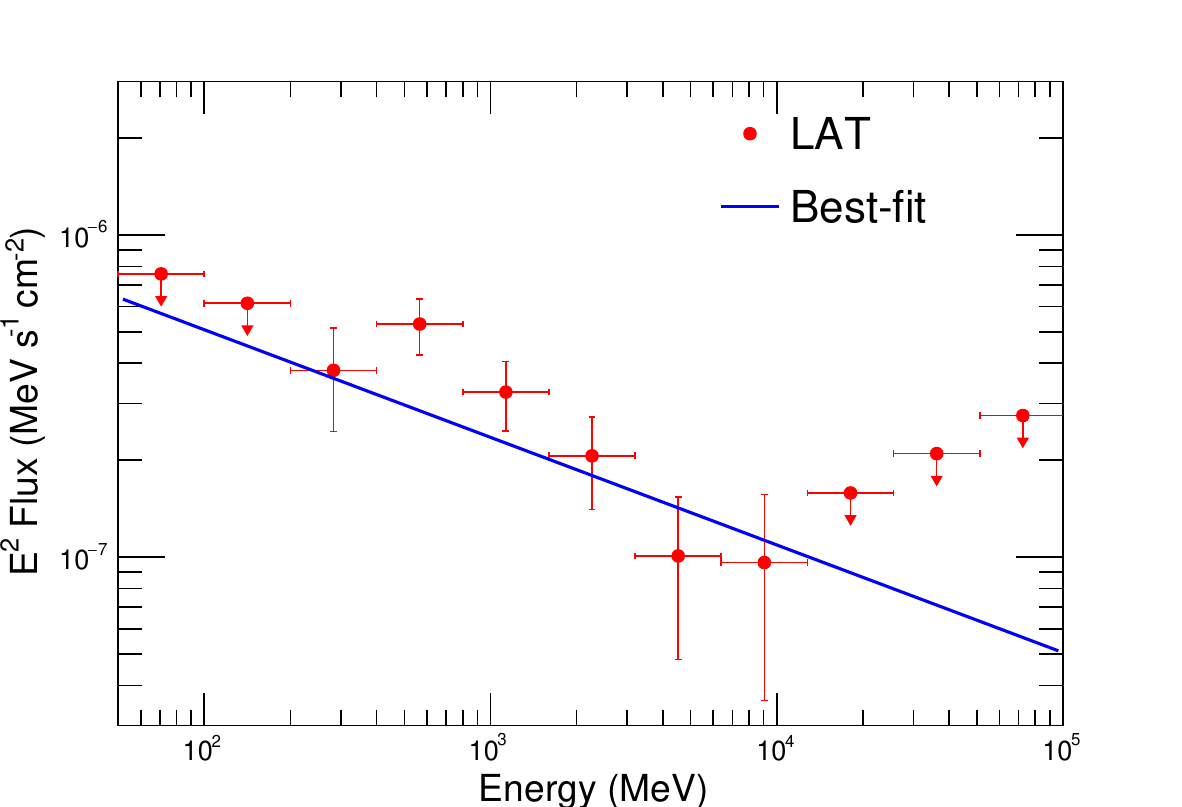}
\put(11,65){(a) Central source}
\end{overpic}
\end{minipage}
\begin{minipage}{0.5\textwidth}
\centering
\begin{overpic}[width=\textwidth]{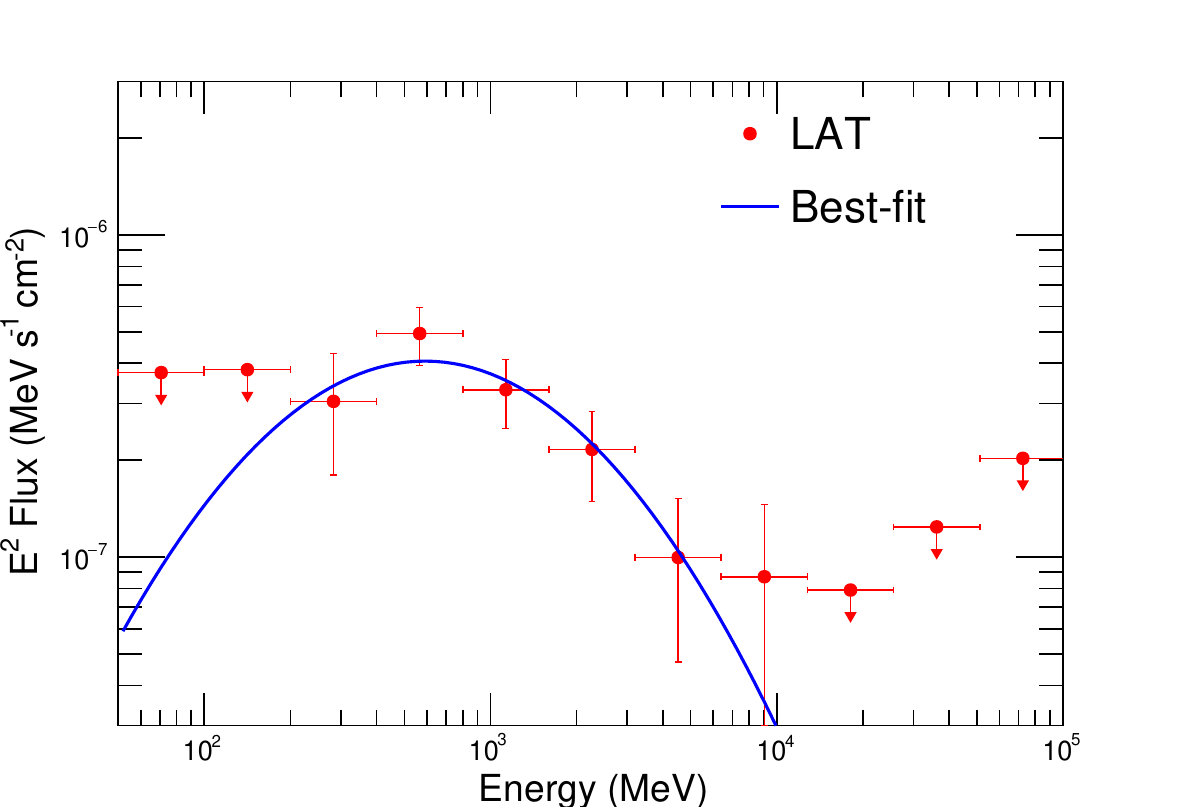}
\put(11,65){(b) Central source}
\end{overpic}
\end{minipage} 
\\
\\
\begin{minipage}{0.5\textwidth}
\centering
\begin{overpic}[width=\textwidth]{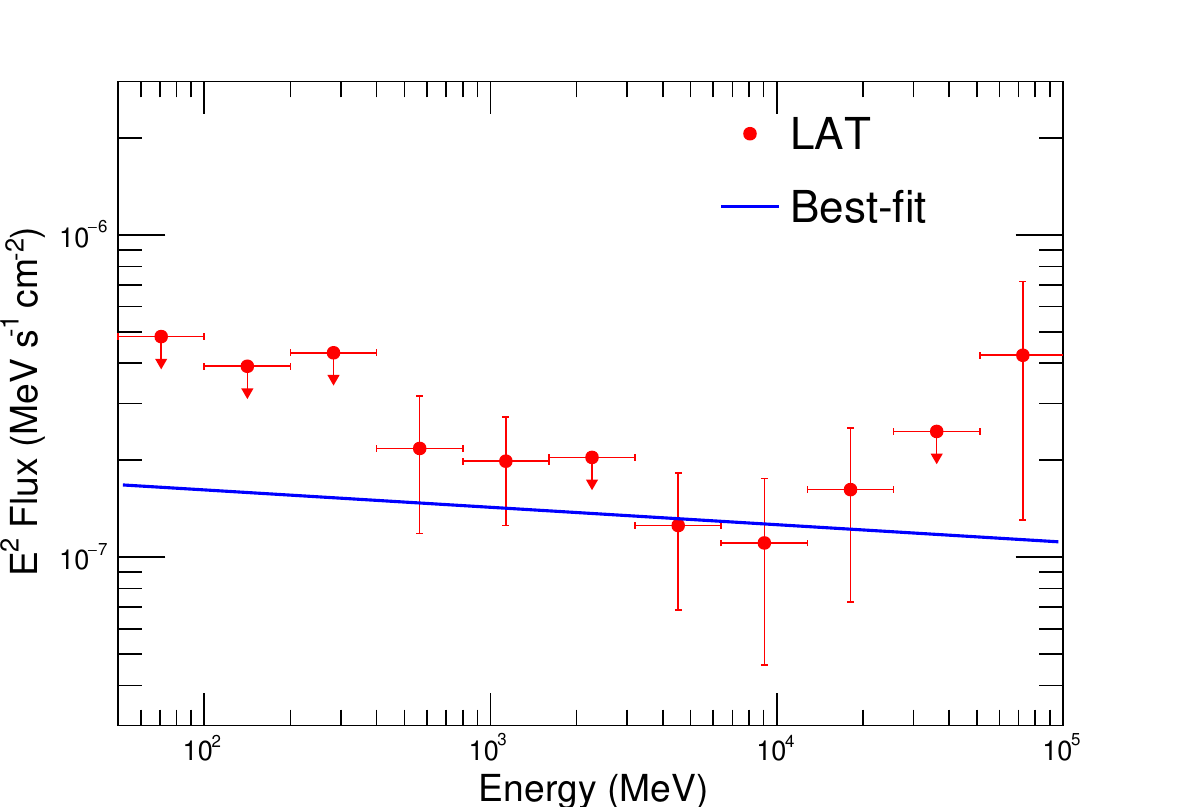}
\put(11,65){(c) SE}
\end{overpic}
\end{minipage}
\begin{minipage}{0.5\textwidth}
\centering
\begin{overpic}[width=\textwidth]{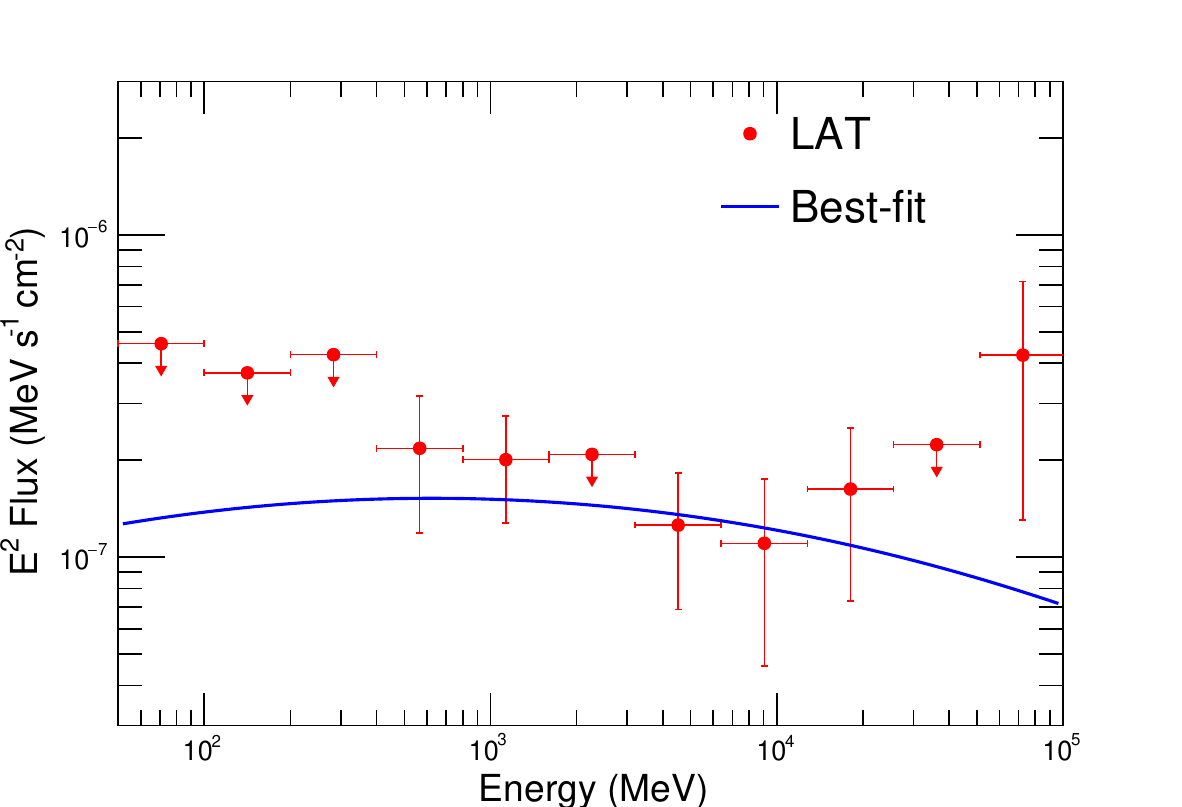}
\put(11,65){(d) SE}
\end{overpic}
\end{minipage}
\end{tabular}
\\
\caption{
SEDs and best-fit spectral models of M31's central source (panels a and b) and the SE source (panels c and d). The results from the PL model are shown in panels a and c, and those from the logP model are shown in panels b and d.
{Alt text: Four SED plots for M31.}
}
\end{figure}

\begin{figure}[htbp]
\begin{tabular}{cc}
\begin{minipage}{1\textwidth}
\centering
\begin{overpic}[width=\textwidth]{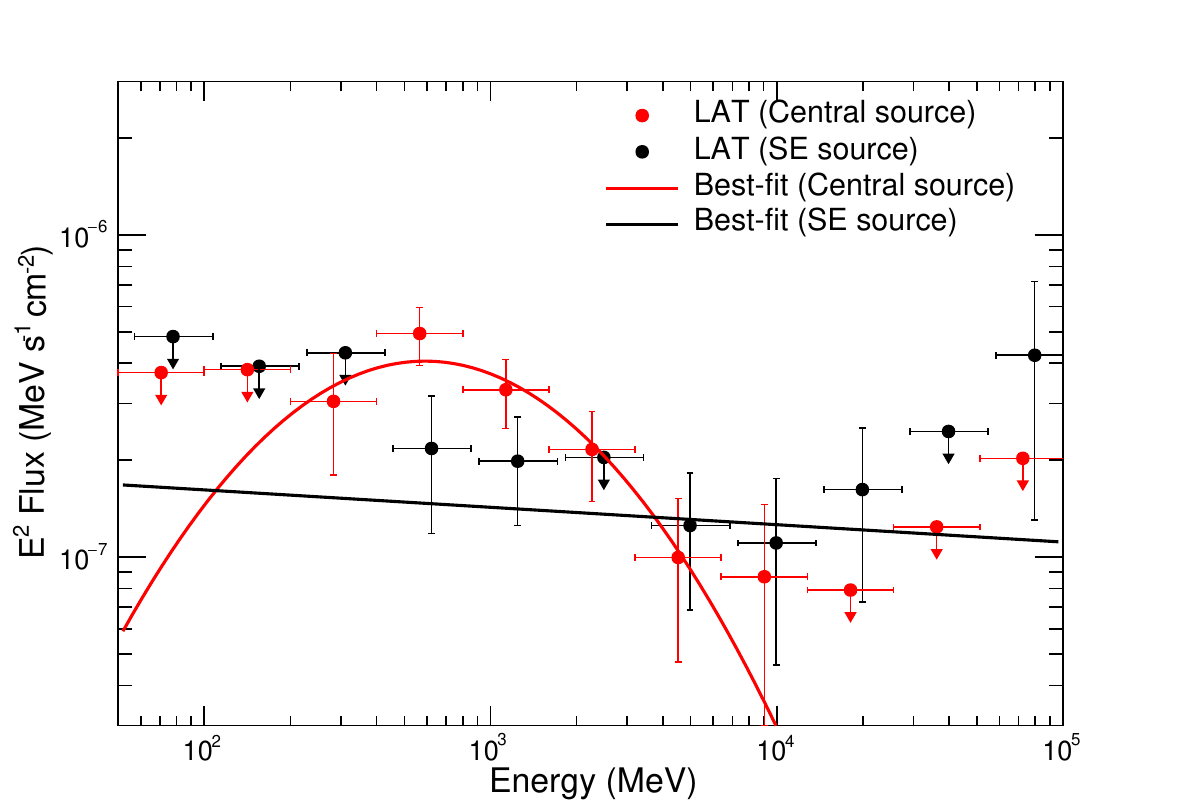}
\end{overpic}
\end{minipage} 
\end{tabular}
\\
\caption{SED and best-fit spectral models for the M31's central source (in red, logP model is used) and the SE source (in black, PL model is used). For clarity, 
SED data points for the SE source are shifted horizontally.
{Alt text: An SED plot for M31.}}
\end{figure}

\vspace{0.5cm}
\begin{figure}[htbp]
\begin{tabular}{cc}
\begin{minipage}{1.0\textwidth}
\centering
\begin{overpic}[width=\textwidth]{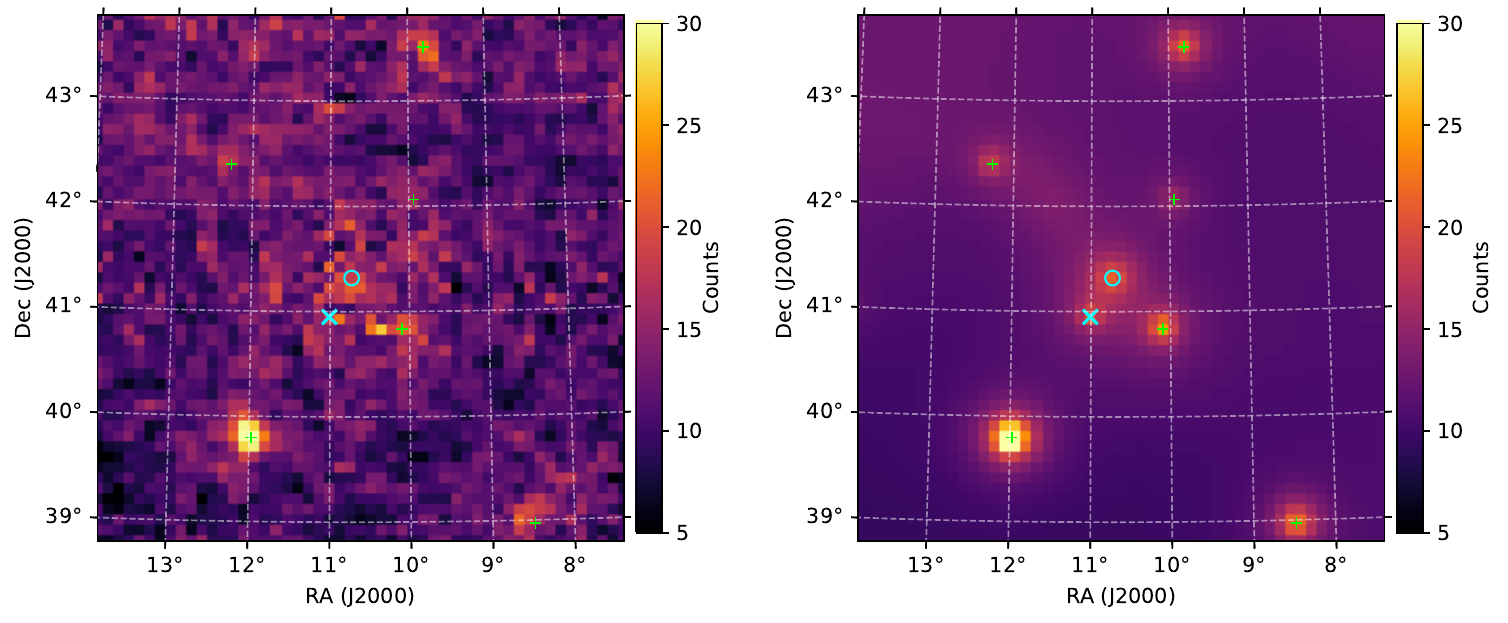}
\put(7,42){(a) M31 ($\ge$400~MeV)}
\put(57,42){(b) M31 ($\ge$400~MeV)}
\end{overpic}
\end{minipage}
\vspace{0.5cm}
\end{tabular}
\caption{(a) Data cont map and (b) model count map in the vicinity of M31 
in {$\ge$}400~MeV. 
In both panels, the position of the M31's central source
is indicated by a cyan circle, that of the SE source is indicated by a cyan cross, 
and those of 4FGL sources are indicated by green crosses.
{Alt text: Two data count maps and two model count maps for the M31 region.}
}
\end{figure}

\clearpage
\subsection{Other Objects}
For other weak sources where photon statistics are limited
(Arp~220, NGC~2146, Arp~299, and M33; each has TS of {$\sim$}50), we adopted the PL model for simplicity.
Use of the logP model reduces the flux by 15\% for M33 and ${\le}$7\% for other sources and will not affect the discussion in Section~4.
Regarding M33, \citet{FermiM33} reported that the $\gamma$-ray emission is
located at NGC~604, an H\,\emissiontype{II} region situated ${\sim}0\fdg14$ northeast of the galaxy's center. We confirmed their finding and modeled the $\gamma$-ray emission as a point source at
NGC~604.
The resulting spectral parameters and numerical values of the SEDs of four galaxies are listed in Tables~2 and 4, and the spectra are shown in Figure~9;
a wider energy range was used to accommodate their low fluxes.
The data and model count maps are presented in Figure~10.
For reference, the spectra obtained using the logP model are shown in Appendix~1.

Compared to the previous analysis by \citet{Ajello2020},
we obtained softer spectra ($\Delta \Gamma \ge 0.25$) for Arp~220 and
NGC~2146. This difference is attributed to the improved background models and photon statistics.
Contamination from diffuse emission and nearby sources is considerable for Arp~220,
while diffuse emission contamination is severe for NGC~2146 (see Figure~10).
Consequently, significantly reduced $\gamma$-ray fluxes (by a factor of about 2) were obtained
for these galaxies.

The photon index ($\Gamma$) for M33 is also smaller (by about 0.20) than that reported by \citet{Ajello2020},
partially due to the slight offset of the source from the galaxy center. When the source position was fixed at the galaxy center, the resulting index was $2.30 \pm 0.12$, consistent with that ($\Gamma=2.41$)
of \citet{Ajello2020}.

\begin{table}[htbp]
  \tbl{Best-fit spectral parameters of our SFG/SBG samples}{%
  \begin{tabular}{ccccccc}
    \hline
Name & TS & $f_\mathrm{e, \mbox{\scriptsize{0.1--100}}}$ & $\Gamma$ (PL) & $\alpha$ (logP) & $\beta$ (logP) & Eb (logP) \\ 
     &    & $\mathrm{erg~s^{-1}~cm^{-2}}$ & & & & MeV (fix)\\
\hline
M82      & 1702 & 8.19e-12 & -- & $2.021\pm0.046$ & $0.110\pm0.024$ &  956 \\
NGC~253  &  899 & 6.31e-12 & -- & $1.991\pm0.064$ & $0.130\pm0.035$ & 1048 \\
M31(center) &  75.1 & 1.87e-12 & -- & $2.29\pm0.19$ & $0.33\pm0.13$ & 913.1 \\
M31(SE)     &  44.2 & 1.50e-12 & $2.05\pm0.18$ & -- & -- & -- \\
Arp~220  & 53.9 & 1.72e-12 & $2.19\pm0.16$ & -- & -- & -- \\
NGC~2146 & 53.1 & 1.33e-12 & $2.00\pm0.13$ & -- & -- & -- \\
Arp~299  & 55.8 & 1.31e-12 & $2.16\pm0.16$ & -- & -- & -- \\
M33      & 44.6 & 1.58e-12 & $2.20\pm0.12$ & -- & -- & -- \\
  \hline
  \end{tabular}}\label{tab:first}
\begin{tabnote}
Notes: $f_\mathrm{e, \mbox{\scriptsize{0.1--100}}}$ gives the integrated energy flux from 100~MeV to 100~GeV. 
$\Gamma$ is the index of the PL model, and $\alpha$, $\beta$, and $E_\mathrm{b}$ are parameters of the logP model. 
\end{tabnote}
\end{table}

\begin{figure}[htbp]
\begin{tabular}{cc}
\begin{minipage}{0.5\textwidth}
\centering
\begin{overpic}[width=\textwidth]{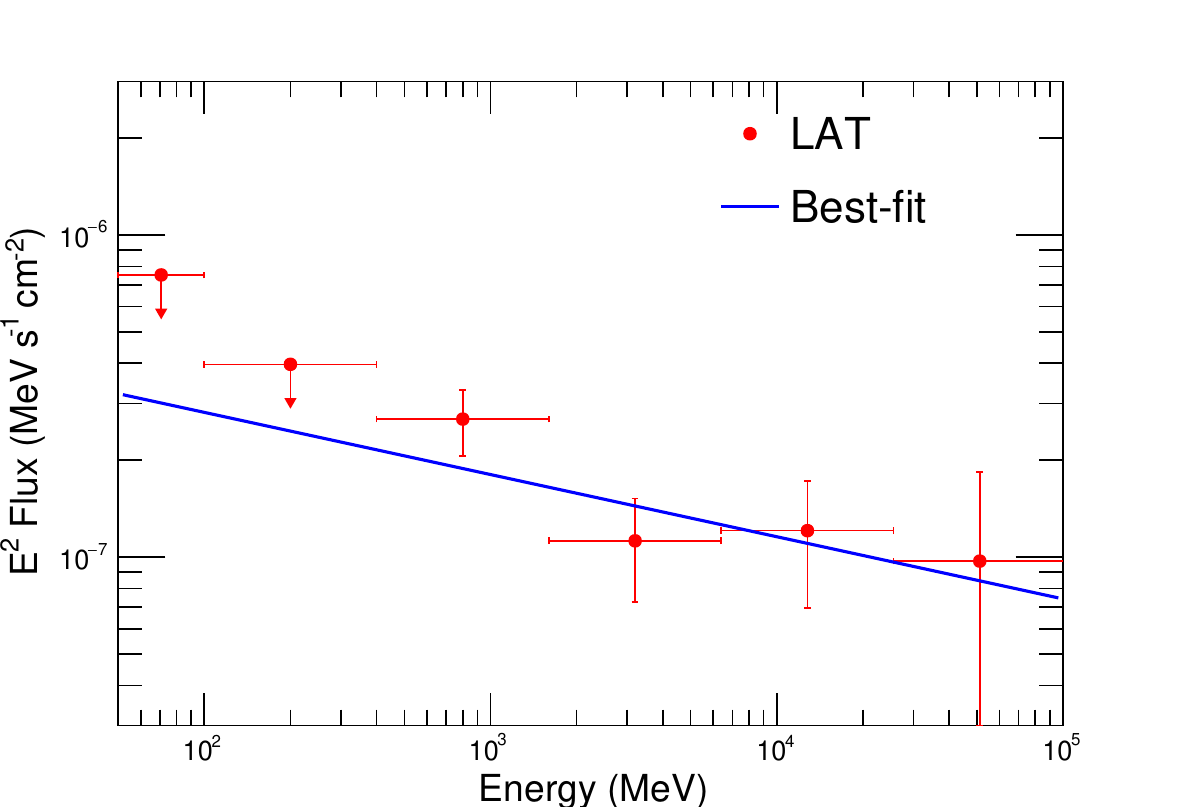}
\put(11,65){(a) Arp~220}
\end{overpic}
\end{minipage}
\begin{minipage}{0.5\textwidth}
\centering
\begin{overpic}[width=\textwidth]{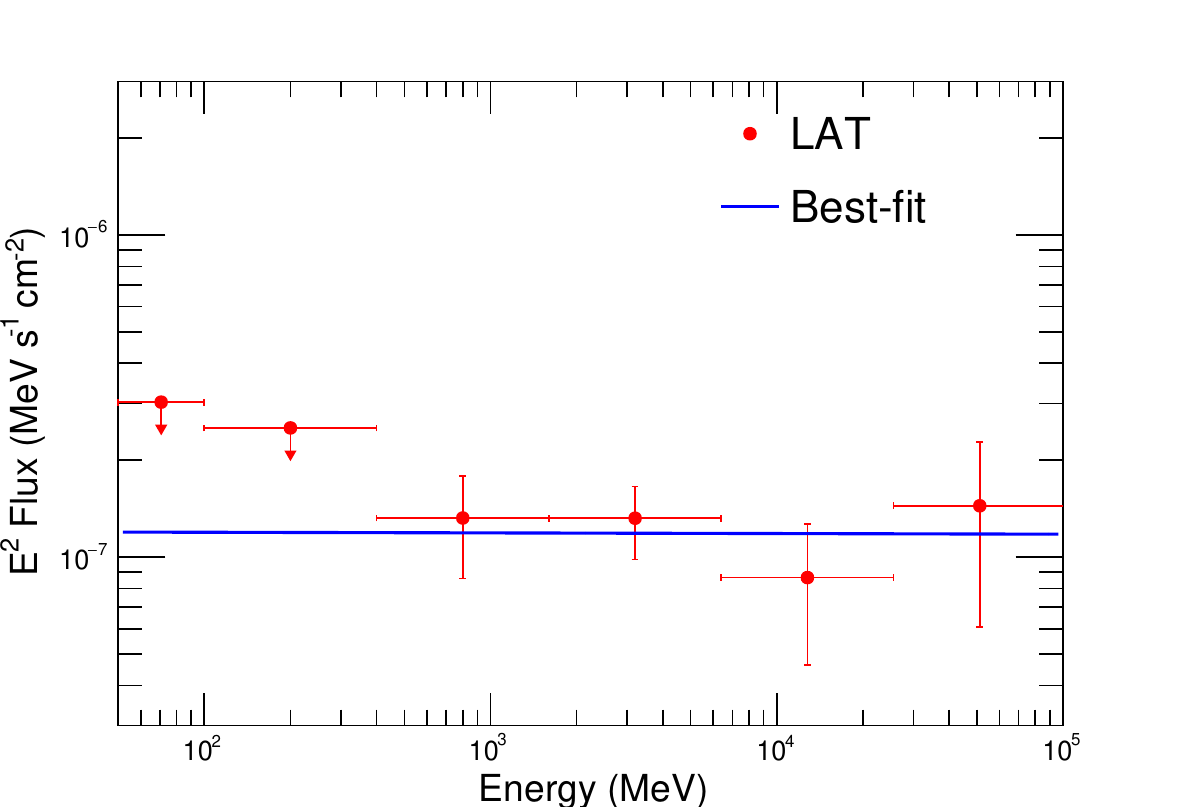}
\put(11,65){(b) NGC~2146}
\end{overpic}
\end{minipage} 
\\
\\
\begin{minipage}{0.5\textwidth}
\centering
\begin{overpic}[width=\textwidth]{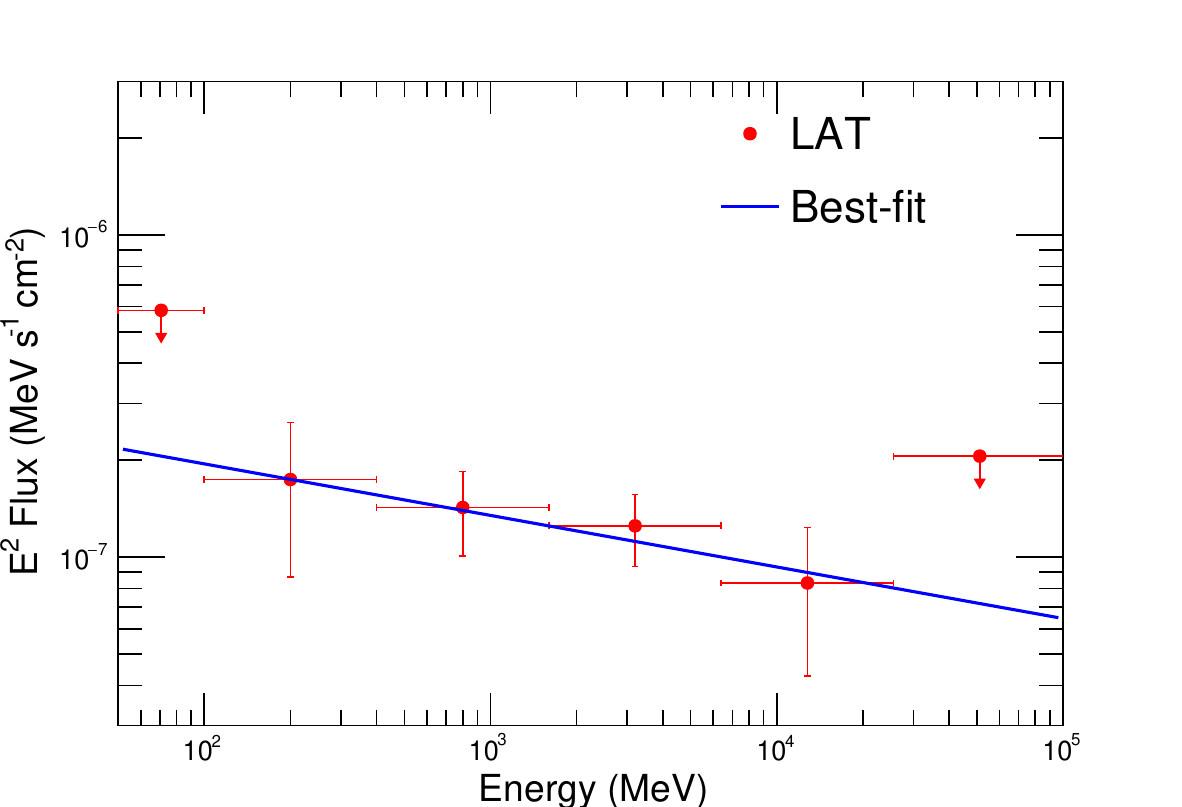}
\put(11,65){(c) Arp~299}
\end{overpic}
\end{minipage}
\begin{minipage}{0.5\textwidth}
\centering
\begin{overpic}[width=\textwidth]{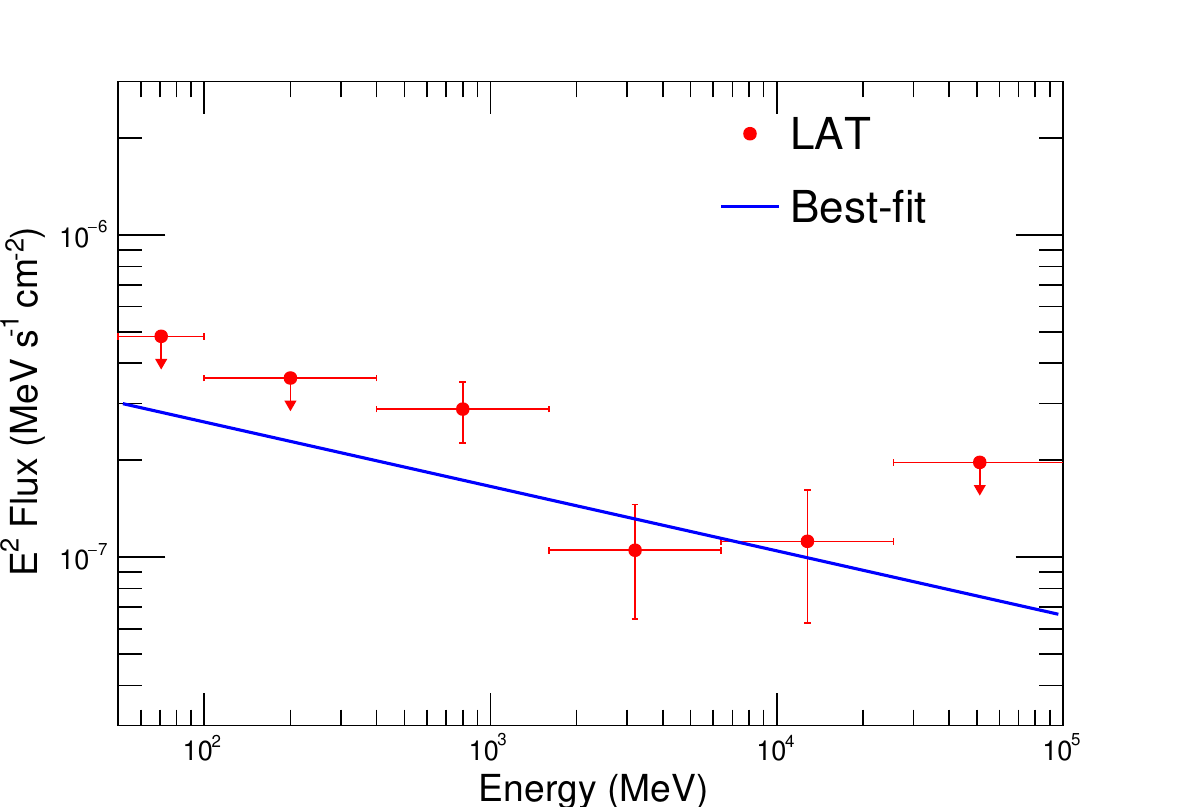}
\put(11,65){(d) M33}
\end{overpic}
\end{minipage}
\end{tabular}
\\
\caption{SED and a best-fit spectral model of (a) Arp~220, (b) NGC~2146,
(c) Arp~299, and (d) M33, obtained by the PL model.
{Alt text: Four SED plots for other objects.}
}
\end{figure}

\vspace{0.5cm}
\begin{figure}[htbp]
\begin{tabular}{cc}
\begin{minipage}{1.0\textwidth}
\centering
\begin{overpic}[width=0.8\textwidth]{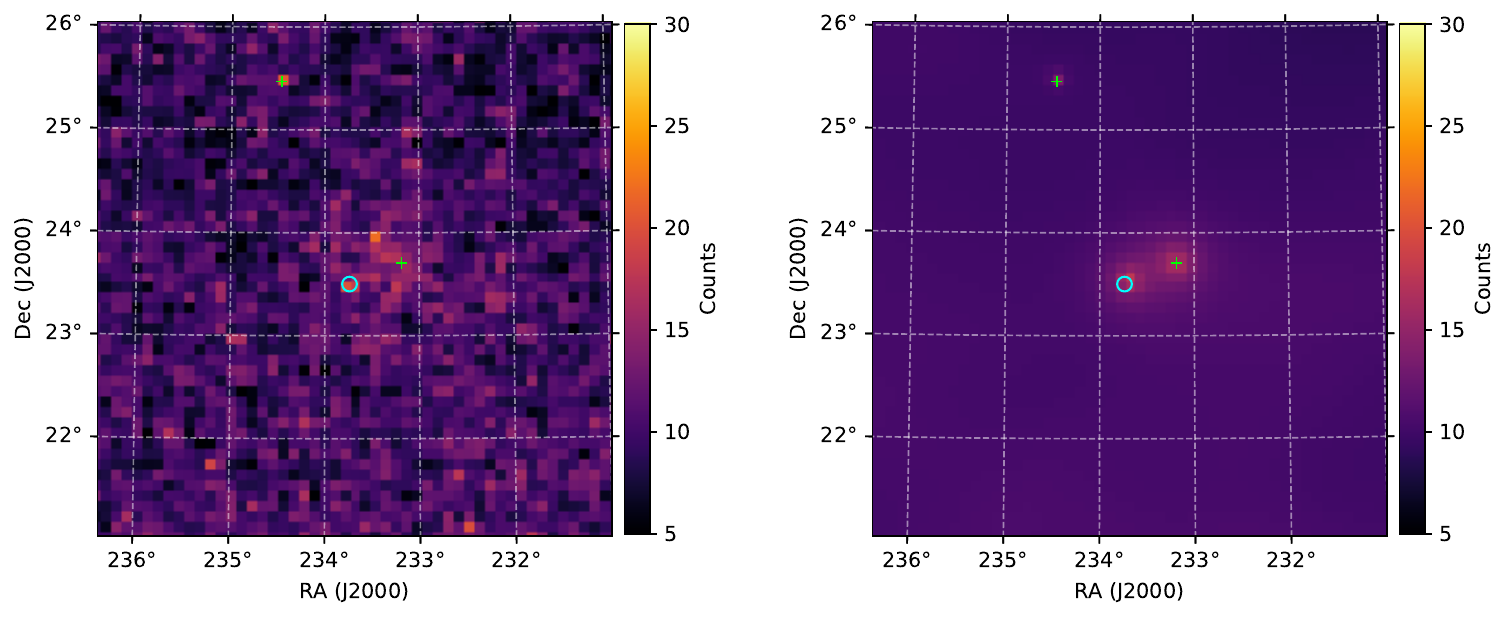}
\put(7,40){(a) Arp220 ($\ge$400~MeV)}
\put(57,40){(b) Arp220 ($\ge$400~MeV)}
\end{overpic}
\end{minipage} 
\\
\\
\begin{minipage}{1.0\textwidth}
\centering
\begin{overpic}[width=0.8\textwidth]{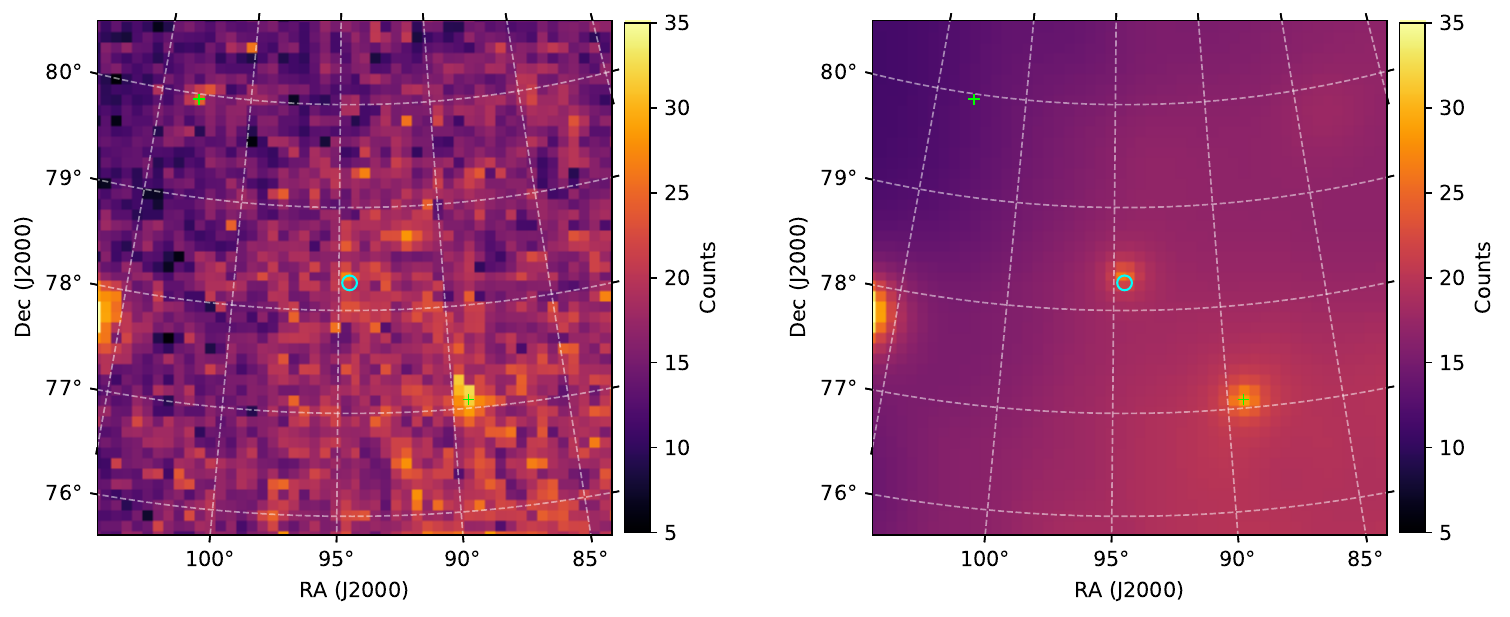}
\put(7,40){(c) NGC~2146 ($\ge$400~MeV)}
\put(57,40){(d) NGC~2146 ($\ge$400~MeV)}
\end{overpic}
\end{minipage} 
\\
\\
\begin{minipage}{1.0\textwidth}
\centering
\begin{overpic}[width=0.8\textwidth]{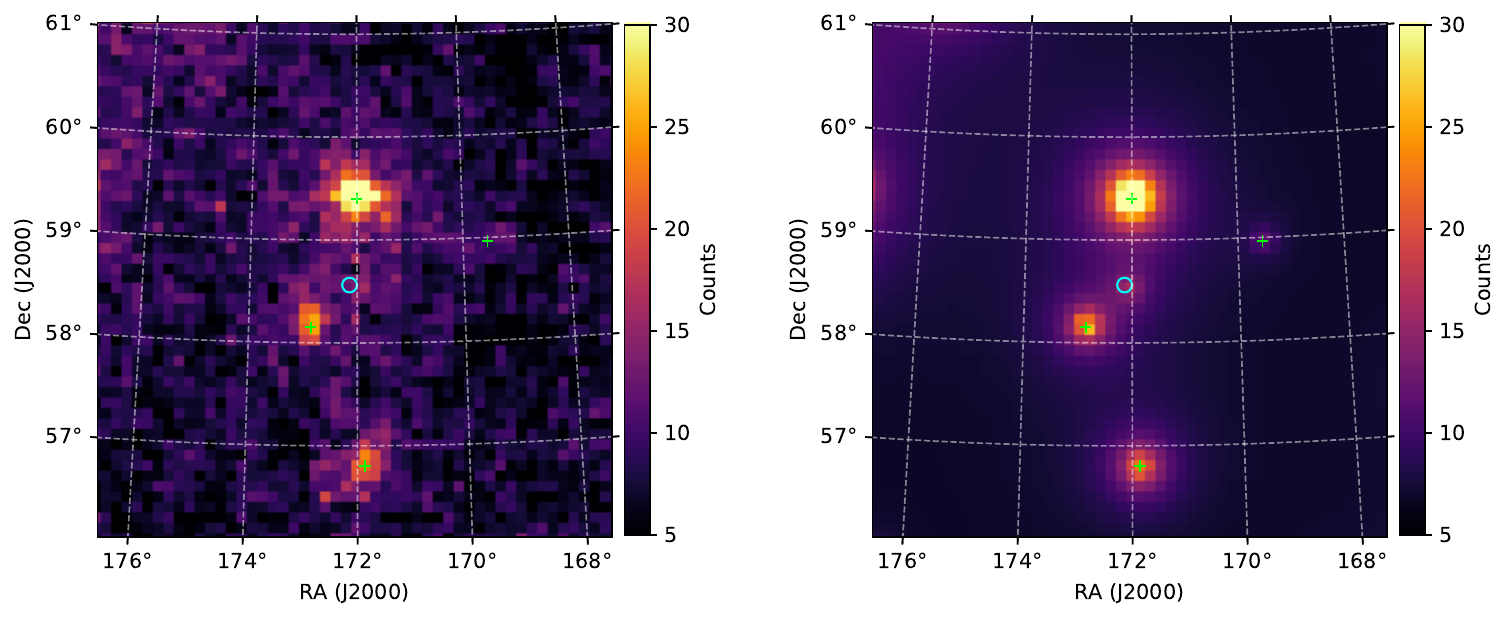}
\put(7,40){(e) Arp299 ($\ge$400~MeV)}
\put(57,40){(f) Arp299 ($\ge$400~MeV)}
\end{overpic}
\end{minipage} 
\\
\\
\begin{minipage}{1.0\textwidth}
\centering
\begin{overpic}[width=0.8\textwidth]{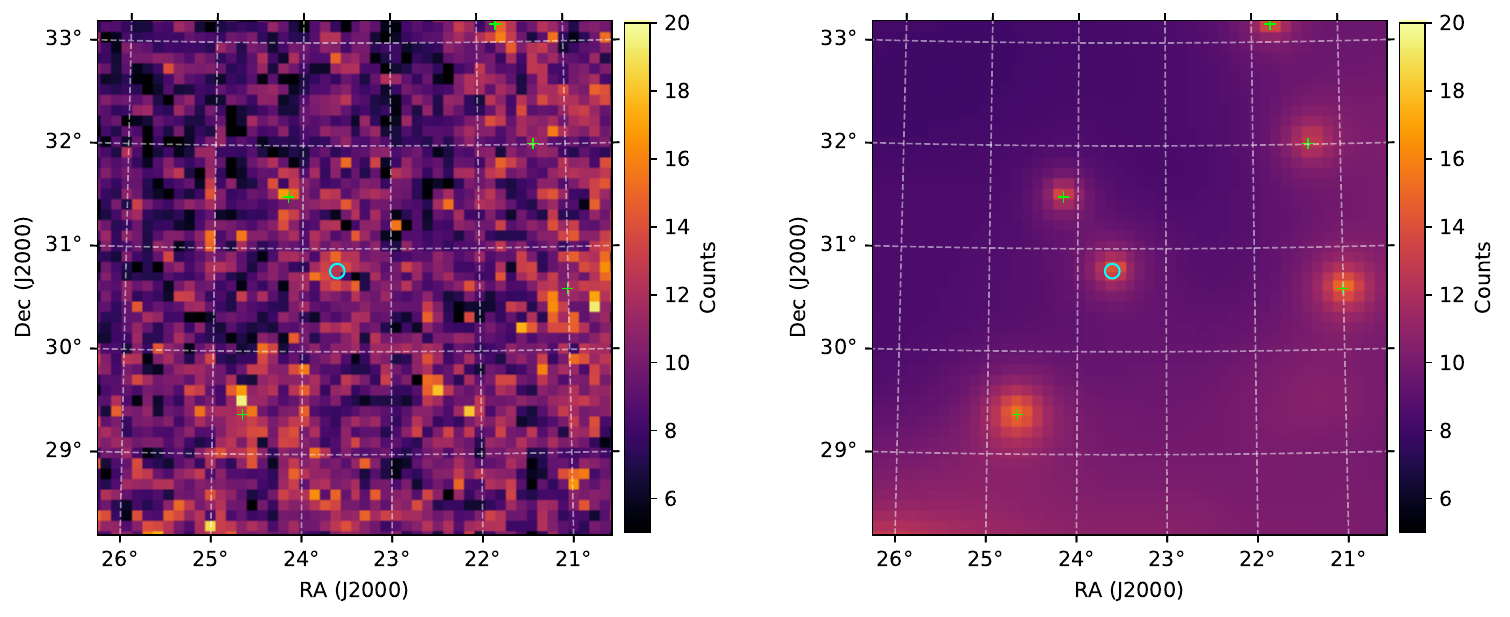}
\put(7,40){(g) M33 ($\ge$400~MeV)}
\put(57,40){(h) M33 ($\ge$400~MeV)}
\end{overpic}
\end{minipage} 
\end{tabular}
\\
\caption{Data count map (left) and model count map (right) in {$\ge$}400~MeV in the vicinity of Arp~220 (panels a and b), NGC~2146 (c and d), Arp~299 (e and f) , and M33 (g and h).
In all panels, the position of the target galaxy is indicated by a cyan circle, and those of 4FGL sources are indicated by green crosses.
{Alt text: Four data count maps and Four model count maps for other object regions.}
}
\end{figure}

\begin{table}[htbp]
  \tbl{SED of bright objects 
  }{%
  \begin{tabular}{ccccc}
    \hline
Energy & M82 (logP) & NGC 253 (logP) & M31 (center; logP) & M31 (SE; PL) \\ 
\cline{2-5}
GeV    & \multicolumn{4}{c}{$\mathrm{10^{-6}~MeV~s^{-1}~cm^{-2}}$} \\
\hline
0.05--0.1 & {$\le$}0.96 & {$\le$}0.63 & {$\le$}0.37 & {$\le$}0.48\\
0.1--0.2 & $0.83\pm0.15$ & $0.52\pm0.16$ & {$\le$}0.38  & {$\le$}0.39\\
0.2--0.4 & $0.96\pm0.11$ & $0.71\pm0.12$ & $0.31\pm0.12$& {$\le$}0.43\\
0.4--0.8 & $1.120\pm0.087$ & $0.872\pm0.094$ & $0.49\pm0.10$ & $0.22\pm0.10$\\
0.8--1.6 & $1.098\pm0.076$ & $0.905\pm0.084$ & $0.331\pm0.080$& $0.199\pm0.073$\\
1.6--3.2 & $1.198\pm0.085$ & $0.824\pm0.089$ & $0.216\pm0.067$& {$\le$}0.20\\
3.2--6.4 & $0.748\pm0.086$ & $0.80\pm0.11$ & $0.100\pm0.052$& $0.125\pm0.057$\\
6.4--12.8 & $0.525\pm0.098$ & $0.42\pm0.11$ & $0.087\pm0.059$& $0.111\pm0.065$\\
12.8--25.6 & $0.57\pm0.14$ & $0.27\pm0.13$ & {$\le$}0.079& $0.162\pm0.089$ \\
25.6--51.2 & $0.22\pm0.12$ & $0.36\pm0.19$ & {$\le$}0.12& {$\le$}0.25\\
51.2--102.4 & $0.15\pm0.15$ & $0.17\pm0.17$ & {$\le$}0.20& $0.42\pm0.29$\\
    \hline
  \end{tabular}}\label{tab:first}
\begin{tabnote}
\end{tabnote}
\end{table}

\begin{table}[htbp]
  \tbl{SED of weak objects, obtained by the PL model. 
  }{%
  \begin{tabular}{ccccc}
    \hline
Energy & Arp 220 & Arp 299 & NGC 2146 & M33 \\ 
\cline{2-5}
GeV    & \multicolumn{4}{c}{$\mathrm{10^{-6}MeV~s^{-1}~cm^{-2}}$} \\
\hline
0.05--0.1   & {$\le$}0.752 & {$\le$}0.584 & {$\le$}0.303 & {$\le$}0.485 \\
0.1--0.4    & {$\le$}0.397 & $0.174\pm0.087$ & {$\le$}0.252 & {$\le$}0.360 \\
0.4--1.6    & $0.268\pm0.062$ & $0.143\pm0.042$ & $0.132\pm0.046$ & $0.288\pm0.062$ \\
1.6--6.4    & $0.112\pm0.040$ & $0.125\pm0.032$ & $0.132\pm0.034$ & $0.105\pm0.041$ \\
6.4--25.6   & $0.121\pm0.051$ & $0.083\pm0.040$ & $0.086\pm0.040$ & $0.112\pm0.050$ \\
25.6--102.4 & $0.097\pm0.087$ & {$\le$}0.206 & $0.144\pm0.084$ & {$\le$}0.197 \\
    \hline
  \end{tabular}}\label{tab:first}
\begin{tabnote}
\end{tabnote}
\end{table}

\clearpage
\section{Discussion}

In the following, we first model the SEDs of two bright SBGs (M82 and NGC~253) and one moderately bright SFG (M31) in Sections 4.1 and 4.2, respectively.
Then we discuss all our samples (including other weak sources) 
collectively in Section~4.3.

\subsection{SED Modeling of $\gamma$-ray bright SBGs}
M82 and NGC~253 are the two nearest starburst galaxies,
and there are several past studies of non-thermal emission from them 
(\cite{M82model2009}, \cite{Peretti2019}, \cite{Persic2025} and references therein); each of these works has its own approach
to achieve their objective.
In order to model the SEDs of 
M82 and NGC~253, 
we employed Naima \citep{Naima}, a Python-based package for computing non-thermal radiation,
including radio, X-ray, GeV and TeV $\gamma$ rays, from relativistic protons and electrons. 
Naima has been widely used in GeV and TeV $\gamma$-ray data analysis. It allows deriving CR proton and electron spectra by fitting multiwavelength data under a single-emission-region approximation.
We fixed interstellar environmental parameters using values reported in the literature for each galaxy, as detailed in Sections~4.1.1 and 4.1.2.
When literature values depend on distance,
such as luminosities, we scaled them to match the adopted distances 
(hereafter $D$) summarized in Table~1.
We then derived the
CR proton and electron spectra by fitting our GeV data together with radio and TeV data
compiled from previous studies.
Diffuse X-ray emission in SBGs may also originate from 
interstellar electrons through synchrotron radiation or IC scattering (e.g., \cite{Lacki2013}). 
However, accurately removing point-source contributions and thermal components is challenging.
Consequently, we excluded X-ray data from the SED fitting, but confirmed that the apparently diffuse X-ray fluxes reported in the literature are consistent with our SED models.

In our SED fitting, we specified three interstellar quantities:
the target gas density for computing pion-decay and electron non-thermal Bremsstrahlung emission;
the ISRF (temperature and energy density) for IC scattering;
and the magnetic field strength for synchrotron emission. 
The CR proton spectrum was modeled as a power-law in momentum.
The injection spectrum of accelerated CR electrons is also expected to follow a power-law, but electrons experience substantial energy losses due to ionization, Bremsstrahlung, synchrotron radiation, and IC scattering.
The total energy-loss rate depends on electron energy $E_\mathrm{e}$ and can be expressed as 
$b(E_\mathrm{e}) = C_{1} + C_{2}E_\mathrm{e} + C_{3}{E_\mathrm{e}}^{2}$. The ratio $C_{1}/C_{2}$ is set by the balance between ionization and Bremsstrahlung losses, while $C_{2}/C_{3}$ is determined by the relative importance of Bremsstrahlung, synchrotron, and IC losses.
Consequently, the steady-state electron spectrum is proportional to the injected spectrum divided by $b(E_\mathrm{e})/E_\mathrm{e}$, resulting in a smoothly-broken double-broken power-law shape
(e.g., \cite{Longair,VERITAS_M82v2}).
The first break energy $E_\mathrm{bk1} = C_\mathrm{1}/C_\mathrm{2}$ is universal,
whereas the second break $E_\mathrm{bk2} = C_\mathrm{2}/C_\mathrm{3}$ depends on the gas density, ISRF energy density, and magnetic-field energy density. 
Thus, our CR spectral model contains four free parameters: 
the normalization and index for CR protons and for CR electrons.

In addition to the synchrotron, free-free emission (thermal Bremsstrahlung) contributes to the radio spectrum. 
According to \citet{Williams2010}, the radio emission of M82 and NGC~253 is severely affected by free-free absorption below {$\sim$}1~GHz and by dust emission above {$\sim$}100~GHz. 
We therefore fitted the radio data in the 1--100~GHz range using a model consisting of
synchrotron plus free-free emission, adopting a fixed spectral index of 0.1 for the free-free component.
Its normalization was allowed to vary freely, giving a total of five
free parameters in the modeling.

\clearpage
\subsubsection{Modeling of M82}
To model M82 SED, we followed \citet{Lacki2013} and related references theirin for interstellar quantities. 
A radius ($R$) of 250~pc and a half-height ($h$) of 100~pc were adopted for the starburst core, based on \citet{Gotz1990}. 
The ISRF was characterized using the SED compilation by \citet{Silva1998}. The peak wavelength $\lambda_\mathrm{p}$ in the infrared band, related to temperature as $\lambda_\mathrm{p}=3.92kT$ 
(where $k$ is the Boltzmann constant and $T$ is temperature), is {$\sim$}60~$\mathrm{\mu m}$, corresponding to a temperature $T_\mathrm{IR} = 60~\mathrm{K}$
\footnote{Although the infrared emission from dust follows a 
modified blackbody, we approximated it as a blackbody to match the input for Naima. As long as IC is not a major component, this does not
affect the resulting CR spectra very much. See also Figures~11 and 12.}. 
The infrared luminosity $L_\mathrm{IR}$ in the wavelength range 8--1000~$\mu$m
was taken from \citet{Sanders2003}\footnote{
$10^{10.77}~L_\odot$ ($=22.5 \times 10^{43}~\mathrm{erg~s^{-1}}$) for $D=3.63$~Mpc.
}.
Considering the starburst core size
 ($2 \pi R^{2} c$, where $c$ is the speed of light),
the energy density in the infrared was estimated as
$U_\mathrm{ph, IR} = 1280~\mathrm{eV~cm^{-3}}$
for $D=3.7$~Mpc.
The peak wavelength and luminosity in the optical band are
about $1/50$ and $1/4$ of those in the infrared, respectively \citep{Silva1998}.
Accordingly, $T_\mathrm{opt} = 3000~\mathrm{K}$ and
$U_\mathrm{ph, opt} = 320~\mathrm{eV~cm^{-3}}$ were obtained. 
In accordance with \citet{Lacki2013}, energy equipartition was assumed, resulting in an estimated magnetic-field strength $B$ of 250~$\mu$G. Given the significant uncertainty of $B$, its influence on the conclusions is addressed in Section~4.3.
For gas number density, we referred to \citet{Weith2001} who measured the molecular mass in the central region of M82
\footnote{$2.3 \times 10^{8}~M_\odot$ for $D=3.9~\mathrm{Mpc}$
},
and derived a proton number density of $n_\mathrm{p} = 220~\mathrm{cm^{-3}}$ for $D=3.7~\mathrm{Mpc}$. 
The interstellar quantities used in the SED fitting are summarized in Table~5.
Using these interstellar parameters and the formulae from \citet{VERITAS_M82v2},
$E_\mathrm{bk1} \sim 0.4~\mathrm{GeV}$
and $E_\mathrm{bk2} \sim 0.7~\mathrm{GeV}$ were obtained and are used as fixed values (Table~6).

We used radio data from \citet{Williams2010},
with the measured radio extent of ${\sim}35\arcsec \times {\sim}10\arcsec$
closely matching the starburst core size. 
The SEDs in the radio,
GeV $\gamma$ rays (Section~3.1), and TeV $\gamma$ rays \citep{VERITAS_M82v2}
were fitted, and the results are summarized in Figure~11. 
The best-fit indices are 2.52 for CR protons and 2.02 for CR electrons, with $K_\mathrm{ep}$ estimated at about $1/30$ at 1~GeV
and about $1/150$ at 10~GeV,
as detailed in Table~6.
In the X-ray regime, the M82 core was examined in detail by \citet{Strickland2007} and \citet{Iwasawa2023} using Chandra data. 
\citet{Iwasawa2023} analyzed 
the starburst core within a region of ${\sim}20\arcsec \times {\sim}10\arcsec$,
removing point sources and the thermal emission contributions, and identified
diffuse hard X-ray emission with a photon index $\Gamma \sim 1.3$.
By converting their reported luminosity\footnote{
(0.7--0.8)$\times 10^{39}~\mathrm{erg~s^{-1}}$ in 4--8~keV for 
$D=3.6$~Mpc
}
into that for $D=3.7~\mathrm{Mpc}$,
we obtained an SED flux ($E^{2} \cdot dN/dE$) of ${\sim}0.70 \times 10^{-12}~\mathrm{erg~s^{-1}~cm^{-2}}$ in 4--8~keV.
We remind that one can resolve only very bright sources, and modeling the unresolved source contribution is uncertain. 
For example, the Galactic Center’s apparently diffuse X-ray emission, once widely considered truly diffuse, was found to be mostly point-source in origin 
(e.g., \cite{Revnivtsev2009}).
The situation is likely more difficult for an external galaxy like M82.
Accordingly, 
we treat this value as an upper limit. We confirm that it
is consistent with our SED model, as illustrated in Figure~11.

\begin{table}[htbp]
  \tbl{Interstellar quantities assumed for the SED fit 
  }{%
  \begin{tabular}{ccccccccc}
    \hline
Name & $R$ & $h$ & $U_\mathrm{ph,IR}$ & $T_\mathrm{IR}$ & $U_\mathrm{ph,opt}$ & $T_\mathrm{opt}$ & $B$ & $n_\mathrm{p}$\\ 
 & pc & pc & $\mathrm{eV~cm^{-3}}$ & K & 
 $\mathrm{eV~cm^{-3}}$ & K & $\mu$G & $\mathrm{cm^{-3}}$\\
\hline
M82      & 250 & 100 &1280 & 60 & 320 & 3000 & 250 & 220\\
NGC~253  &  150 & 50 & 900 & 60 & 225 & 3000 & 210 & 320\\
  \hline
  \end{tabular}}\label{tab:first}
\begin{tabnote}
\end{tabnote}
\end{table}

\begin{table}[htbp]
  \tbl{CR spectral parameters obtained by the SED fit}{%
  \begin{tabular}{ccccccccc}
    \hline
Name & $W_\mathrm{p} (\ge 1~\mathrm{GeV})$ & $\alpha_\mathrm{p}$ & $W_\mathrm{e} (\ge 1~\mathrm{GeV})$ & $\alpha_\mathrm{e}$ & $K_\mathrm{ep, 1GeV}$ & $K_\mathrm{ep, 10GeV}$ & $E_\mathrm{bk1}$ & $E_\mathrm{bk2}$ \\ 
 & $\mathrm{10^{53}~erg}$ &  & $\mathrm{10^{51}~erg}$ &  &  &
& GeV & GeV \\
\hline
M82      & $3.40^{+0.21}_{-0.27}$ & $2.52^{+0.06}_{-0.05}$ & $3.69^{+0.11}_{-0.23}$ & $2.02^{+0.05}_{-0.15}$ & 1/28 & 1/147 & 0.4 & 0.7 \\
NGC~253  & $1.54^{+0.11}_{-0.11}$ & $2.41^{+0.04}_{-0.04}$ & $1.66^{+0.07}_{-0.23}$ & $2.03^{+0.14}_{-0.02}$ & 1/26 & 1/120 & 0.4 & 1.5 \\
  \hline
  \end{tabular}}\label{tab:first}
\begin{tabnote}
Note: $E_\mathrm{bk1}$ and $E_\mathrm{bk2}$ are pre-computed from the assumed interstellar quantities and fixed. 
$W_\mathrm{p}$ and $W_\mathrm{e}$ are best-fit total energies of CR protons
and electrons, respectively.
$\alpha_\mathrm{p}$ and $\alpha_\mathrm{e}$ are best-fit indices of
CR protons and electrons, respectively.
They give $K_\mathrm{ep}$ and values at 1~GeV and 10~GeV are also presented.
Best-fit normalizations of the free-free emission are $0.68\pm0.05$ and $0.67\pm0.06$ times the measured flux for M82 (at 92~GHZ) and NGC~253 (at 94~GHz), respectively. \\ 
\end{tabnote}
\end{table}

\begin{figure}[htbp]
\begin{tabular}{cc}
\begin{minipage}{0.5\textwidth}
\centering
\begin{overpic}[width=1\textwidth]{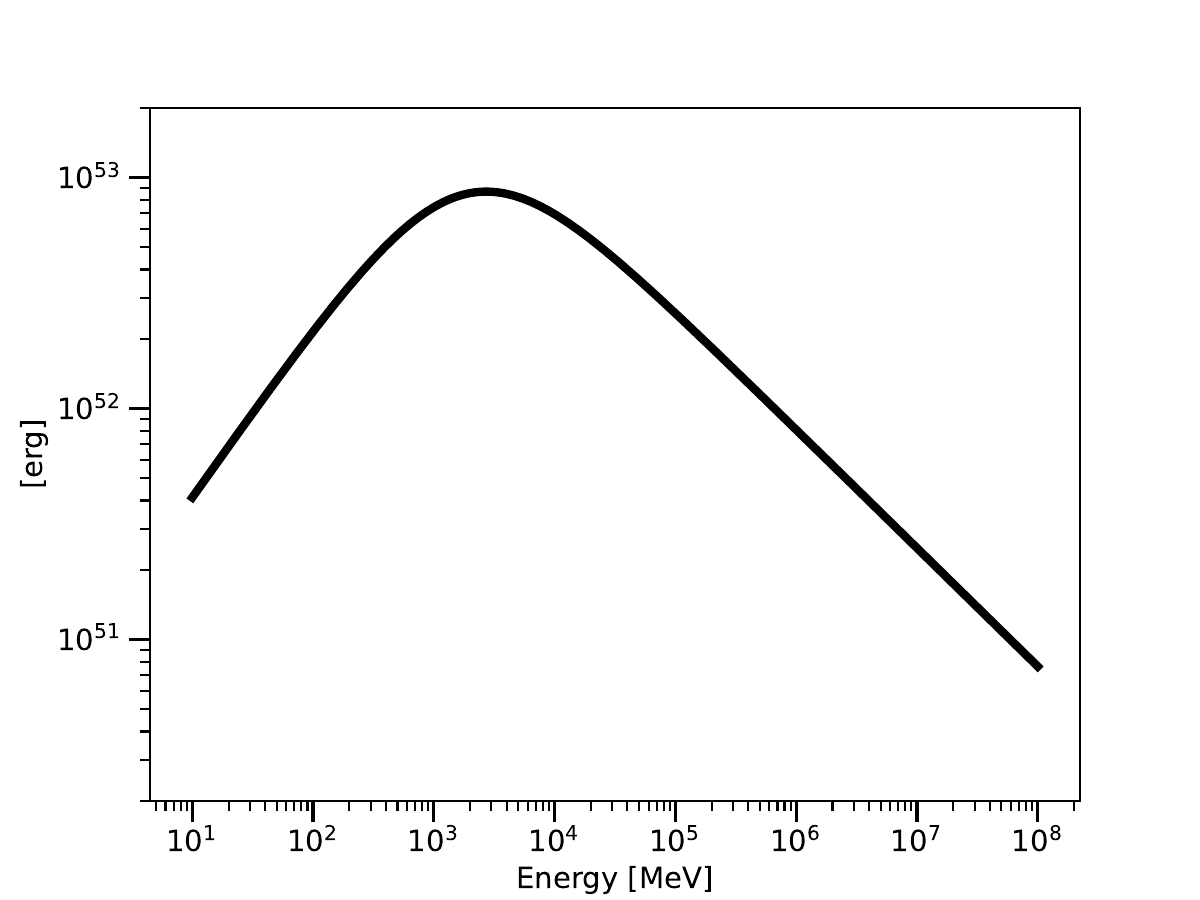}
\put(15,69){(a) proton model spectrum}
\end{overpic}
\end{minipage}
\begin{minipage}{0.5\textwidth}
\centering
\begin{overpic}[width=1\textwidth]{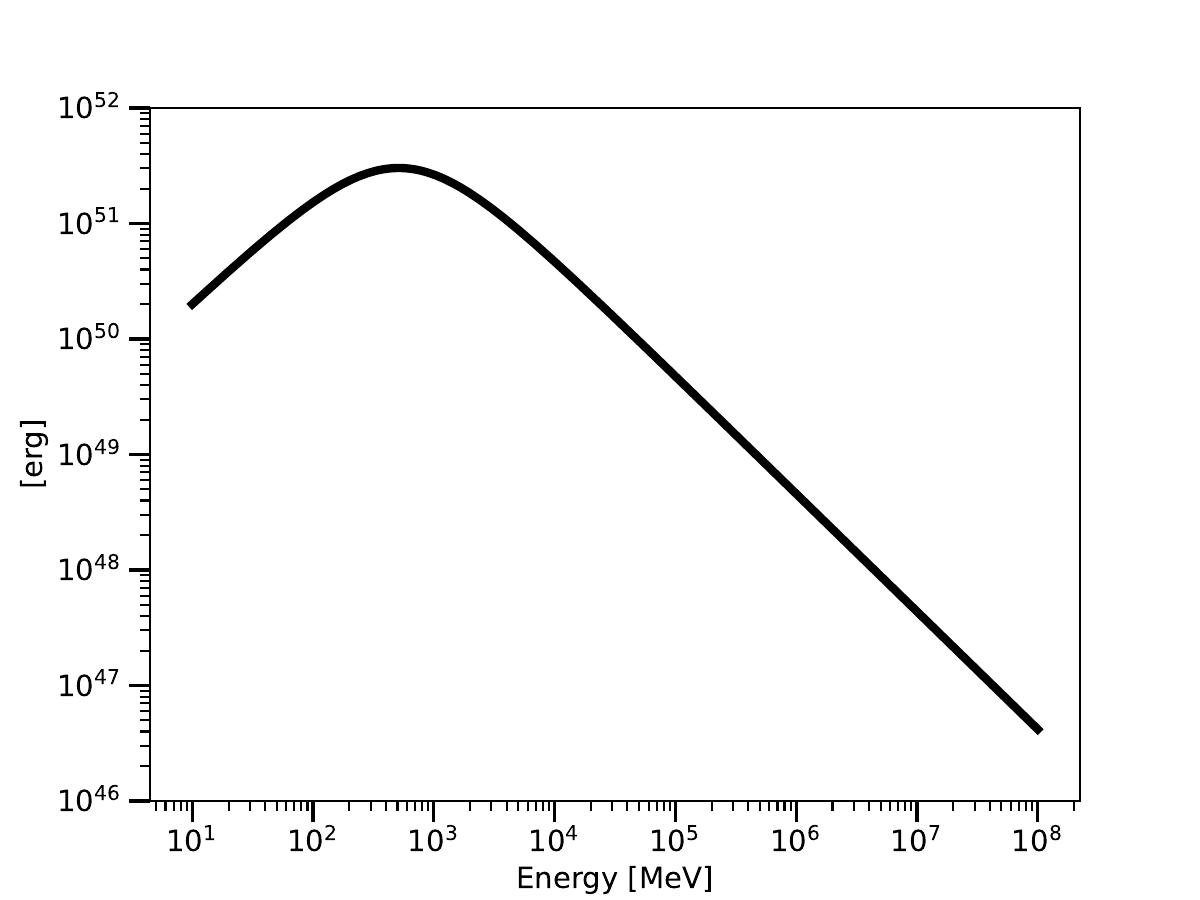}
\put(15,69){(b) electron model spectrum}
\end{overpic}
\end{minipage}
\\
\\
\begin{minipage}{\textwidth}
\centering
\begin{overpic}[width=1\textwidth]{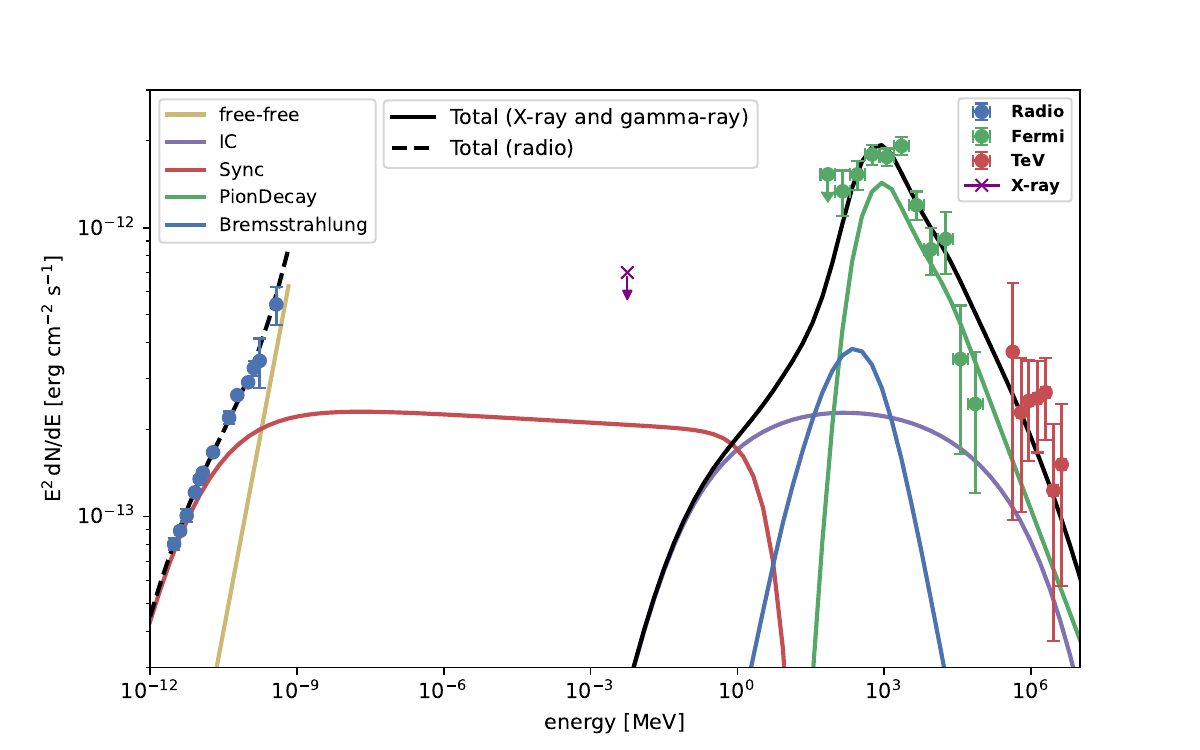}
\put(15,56){(c) SED of M82}
\end{overpic}
\end{minipage} 
\end{tabular}
\caption{(a) Best-fit proton spectrum and (b) electron spectrum for M82. 
(c) SED and best-fit model components for M82. 
{Alt text: Three plots to show CR spectral models and the SED fit for M82.}
}
\end{figure}

\clearpage
\subsubsection{Modeling of NGC~253}
We also followed \citet{Lacki2013} and references therein for the interstellar parameters of NGC~253. For the starburst core, we adopted $R=150~\mathrm{pc}$ and $h=50~\mathrm{pc}$.
Using the infrared luminosity reported by \citet{Sanders2003}\footnote{
$L_\mathrm{IR}$ as $10^{10.29}~L_\odot$ for $D=3.10$~Mpc
},
we obtained $L_\mathrm{IR}$ as $10^{10.49}~L_\odot$ for $D=3.3$~Mpc. 
Approximately half of this luminosity originates from the nuclear starburst \citep{Melo2002},
and we halved $L_\mathrm{IR}$ above.
Following \citet{Lacki2013}, we assumed $U_\mathrm{ph, IR}/U_\mathrm{ph, opt}=4$
and energy equipartition as in M82, 
giving the ISRF and magnetic-field energy densities 
to be
{$\sim$}70\% of those for M82.
Specifically, we adopted $U_\mathrm{ph, IR}=900~\mathrm{eV~cm^{-3}}$, $U_\mathrm{ph, opt}=225~\mathrm{eV~cm^{-3}}$,
and $B=210~\mu$G.
For the ISM gas mass, we referred to \citet{Sakamoto2011},
who identified
several molecular clumps in the central molecular zone
\footnote
{They identified five molecular clumps,
each with ${\sim}10^{7}~M_\odot$,
and estimated the total mass of $6.2 \times 10^{7}~M_\odot$ for $D=3.5$~Mpc}.
A proton number density was estimated as $n_\mathrm{p} = 320~\mathrm{cm^{-3}}$
for $D=3.3$~Mpc. 
Consequently, $E_\mathrm{bk2}$ was set to 1.5~GeV, about twice the value adopted for M82. These parameters are summarized in Tables~5 and 6.

\citet{Williams2010} measured the radio emission at 1-7 GHz and compiled additional data at 20--100~GHz from the literature. They found that the emission is spatially extended and modeled it as the sum of core and extended components.
Their adopted core size of $16\arcsec \times 10\arcsec$ is consistent
with the starburst-core dimensions used in our modeling.
We fitted the SED using the core-only radio data from \citet{Williams2010}, our GeV $\gamma$-ray data (Section~3.2), and TeV $\gamma$-ray data from \citet{HESS_N253v2}.
The resulting fit is shown in Figure~12.
The best-fit spectral indices are 2.41 for CR protons and 2.03 for CR electrons,
with $K_\mathrm{ep}$ at 1~GeV and 10~GeV similar to those for M82. These results are summarized in
Table~6.
For the X-ray band, we referred to \citet{Wik2014},
who analyzed NuSTAR and Chandra data and estimated the IC flux (with $\Gamma \sim 1.6$) from the core using several emission-region models. Their upper limit for a circular region of radius 15\arcsec \  is $2.0 \times 10^{-13}~\mathrm{erg~s^{-1}~cm^{-2}}$ at 20~keV. This limit is consistent with our model prediction, as shown in Figure~12.

\begin{figure}[htbp]
\begin{tabular}{cc}
\begin{minipage}{1\textwidth}
\centering
\begin{overpic}[width=\textwidth]{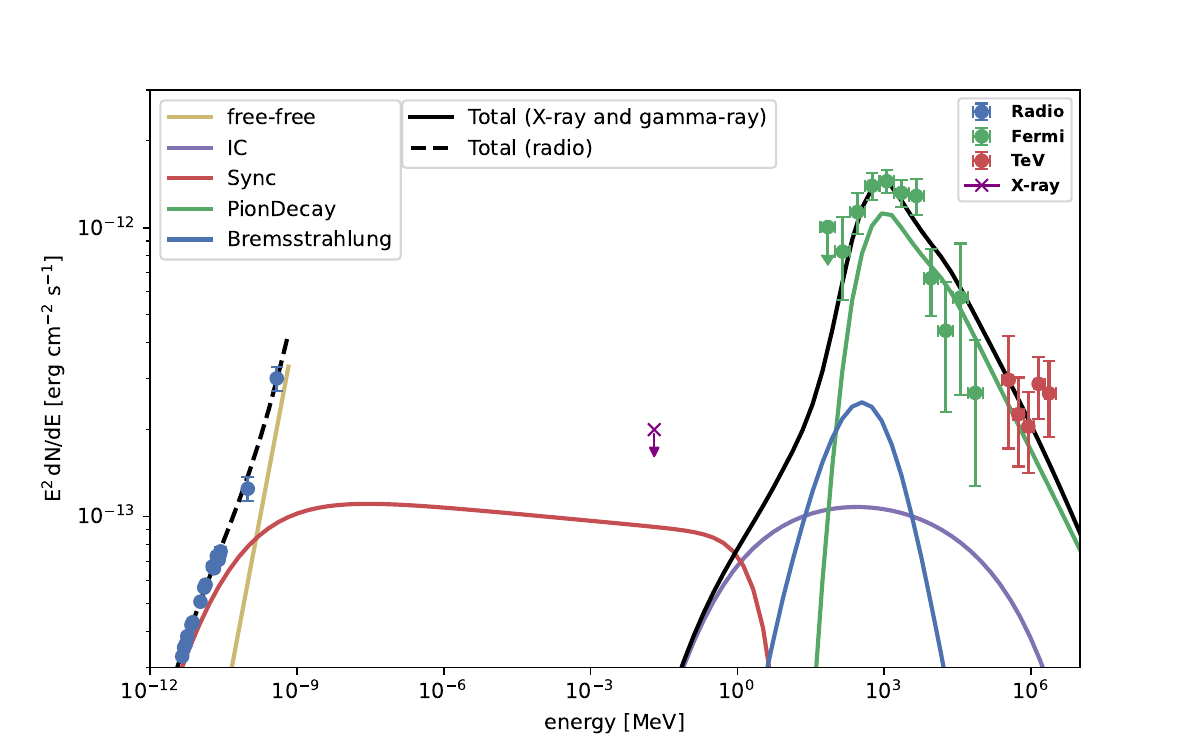}
\put(18, 56){SED of NGC 253}
\end{overpic}
\end{minipage} 
\end{tabular}
\caption{SED and best-fit model components for NGC~253. 
{Alt text: A plot to show the SED fit for NGC~253.}
}
\end{figure}

\subsection{SED Modeling of M31}
As described in \S~3.1, we found that the $\gamma$-ray emission from M31 consists of two components: one located at the galaxy's center
and another ${\sim}0\fdg4$ to the SE, confirming the results of \citet{Xing2023}.
Given the 4FGL source density of ${\sim}0.17~\mathrm{deg^{-2}}$, the probability of a background source lying within $0\fdg4$ is non-negligible, at roughly the 10\% level. 
In the following, we therefore focus on the central source.

The central source dominates the $\gamma$-ray emission around 1~GeV (see Figure~7), 
and its physical origin has been debated (e.g., \cite{FermiM31v2}) due to its apparently compact morphology.
In our analysis, the 95\% upper limit on its radius is only $0\fdg33$,
corresponding to 4.5~kpc --- well below the extent of the 10~kpc ring (e.g., \cite{Smith2012}).
Although millisecond pulsars (MSPs) 
may contribute (e.g., \cite{Persic2024}),
and previous studies have discussed a significant contribution (e.g., \cite{FermiM31v2})

to explain the location and size of the $\gamma$-ray emission,
the spectrum we obtained
appears to contradict
this interpretation. \citet{Calore2014} showed that the spectrum of Galactic MSPs is
well described by an exponential-cutoff power law with a mean photon index of {$\sim$}1.3 and a cutoff energy of {$\sim$}2.4~GeV, producing a characteristic SED peak at 
2--3~GeV (see their Figure~11). This spectral shape is inconsistent with the $\gamma$-ray
spectrum from M31,
where we observe the cutoff at ${\le}$1~GeV (see Figure~7),
requiring a much softer spectrum for MSPs in M31.

Setting aside the small source extension, let us first
investigate the scenario of interstellar $\gamma$-ray emission originating from star-formation activity.
Due to the limited photon statistics, with the flux less than half those 
of M82 and NGC~253, 
we do not fit the SED. Instead, we employed the theoretical modeling of $\gamma$-ray emission from SFGs and SBGs developed by \citet{Chen2025}.
This model estimates $\gamma$-ray emission from CR protons and leptons based on several key galactic parameters: star formation rate $\psi$, gas mass $M_{\mathrm{gas}}$, stellar mass $M_{\mathrm{star}}$, effective radius $R_{\mathrm{eff}}$, and luminosities in infrared and optical bands ($L_{\mathrm{IR}}$ and $L_{\mathrm{opt}}$), under the assumption of
a one-zone disk-like geometry. 
In the original formulation, 
CR particles are assumed to be confined within the gas scale height $H_{\rm gas}$ from the galactic plane. To improve the accuracy of $\gamma$-ray emission predictions,
we have newly introduced a CR scale height $H_{\rm CR}$, which is
significantly larger than $H_{\rm gas}$ in non-calorimetric cases.
We have also updated the key model parameters to reproduce the canonical values of mean path length (grammage) and CR escape time for the Milky Way. 
Further methodological details are provided in Appendix~2. 
We remind readers that this modeling does not fit the observed SED; instead, it determines the $\gamma$-ray spectral shape and flux normalization a priori.
We then compared the model prediction of
the interstellar $\gamma$-ray emission
with the measured spectrum
to interpret the observed $\gamma$-ray emission.

Specifically, we adopted the following parameters and summarized them in Table~7;
$\psi=0.53~\mathrm{M_{\odot}~yr^{-1}}$, $M_\mathrm{gas}=7.6\times10^{9}~M_{\odot}$,
$M_\mathrm{star}=86\times10^{9}~M_{\odot}$ and $R_\mathrm{eff}=10~\mathrm{kpc}$.
These values are adopted from \citet{Shimono2021}, with $R_\mathrm{eff}$ updated to correspond to the gas disk size (10~kpc ring).
$L_{\mathrm{IR}}$ and $L_{\mathrm{opt}}$, which determine the seed photon distribution for IC scattering, are also sourced from the literature and are presented in the Table~7.
Three characteristic timescales (proton-proton collision time $t_\mathrm{pp}$, 
diffusion loss time $t_\mathrm{diff}$, and advection loss time $t_\mathrm{adv}$) calculated by the model are shown in Figure~13(a) as a function of CR proton energy 
$E_\mathrm{p}$. 
For comparison, we also give time scales of the Milky Way. In the GeV (and TeV) energy band,
$t_\mathrm{pp}$ is more than one order of magnitude larger than $t_\mathrm{diff}$, resulting in
a calorimetric efficiency 
as a function of $E_\mathrm{p}$, $f_\mathrm{cal}(E_\mathrm{p})$, less than
10\% (See Figure~13(b)). 
Accordingly,
the predicted flux of the interstellar $\gamma$-ray emission 
is low despite a high star-formation rate, and the calculated spectrum is comparable to or lower than the measurement, as summarized in Figure~13(c).
This explains why we do not detect significant $\gamma$-ray emission from the 10~kpc ring in M31.
Interestingly, predicted $\pi^{0}$-decay $\gamma$-ray emission peaks at {$\sim$}1~GeV and decays steeply below 200~MeV, being compatible with the data.

As discussed above, the scenario in which the central source originates from the interstellar $\gamma$-ray emission 
is problematic given the observed small source extent;
there must be a considerable contribution from unresolved sources located in the vicinity ({$\le$}4.5~kpc) of the M31 center.
Our analysis and modeling indicate that the contribution from unresolved 
sources is considerable,
with the spectral shape
peaking at {$\sim$}1~GeV.
One possible solution is that the average spectrum of MSPs in M31 is much softer than that in the Milky Way.
Indeed, Galactic MSPs exhibit softer spectra when the rotational power is lower (i.e., the PSR is older), as discussed by \citet{Kerr2026}.
Hence, if the average age of MSPs in M31 is much older,
MSPs may explain the observed spectrum of M31's central source.
Although this is an interesting scenario, it requires detailed modeling of the MSP population in M31 and is outside the scope of this work.

The magnetic field strength ($B$) in the model, determined from $\psi$ and the
galaxy size, is {$\sim$}1.7~$\mu$G.
This value is lower than several estimates derived from energy equipartition between CRs and the magnetic field (e.g., \cite{Beck2025}; \cite{Fletcher2004}). Nevertheless, using $B=6$~$\mu$G from these estimates yields similar $\gamma$-ray emission, as shown in Figure~13(c). Therefore, the discussion on the
comparison between the $\gamma$-ray spectral data and the model is not affected.

\begin{table}[htbp]
  \tbl{Parameters to model $\gamma$-ray emission from M31, NGC~2146, 
  and Milky Way 
  }{%
  \begin{tabular}{cccccccc}
    \hline
Name & $\psi$ & ${M_\mathrm{gas}}$ & ${M_\mathrm{star}}$ & ${R_\mathrm{eff}}$ & $D$ & ${L_\mathrm{IR}}$ & ${L_\mathrm{opt}}$\\
 & $\mathrm{M_{\odot}~yr^{-1}}$ & $10^{9}~M_{\odot}$ & $10^{9}~M_{\odot}$ & kpc & Mpc & $10^{43}~\mathrm{erg~s^{-1}}$ & $10^{43}~\mathrm{erg~s^{-1}}$ \\
\hline
M31(center) & 0.53 &  7.6 & 86.0 & 10 & 0.78 & 0.94 & 12.2 \\
NGC~2146    & 11.4 & 10.4 & 87.1 & 1.7 & 17.2 & 45.0 & 7.27\\
Milky Way   & 2.6  &  4.9 & 64.3 & 6 & -- & 5.2 & 19.6 \\
  \hline
  \end{tabular}}\label{tab:first}
\begin{tabnote}
Notes: Parameters for Milky Way and NGC~2146 are taken from \citet{Chen2025}.
For M31, $L_{\mathrm{IR}}$ and $L_{\mathrm{opt}}$ are taken from \citet{Ajello2020} and \citet{deVaucouleurs1991}, respectively.
Other M31's parameters are taken from \citet{Shimono2021}, but
$R_\mathrm{eff}$ is updated to correspond to the gas disk size (see text).
\end{tabnote}
\end{table}

\begin{figure}[htbp]
\begin{tabular}{cc}
\begin{minipage}{0.5\textwidth}
\centering
\begin{overpic}[width=\textwidth]{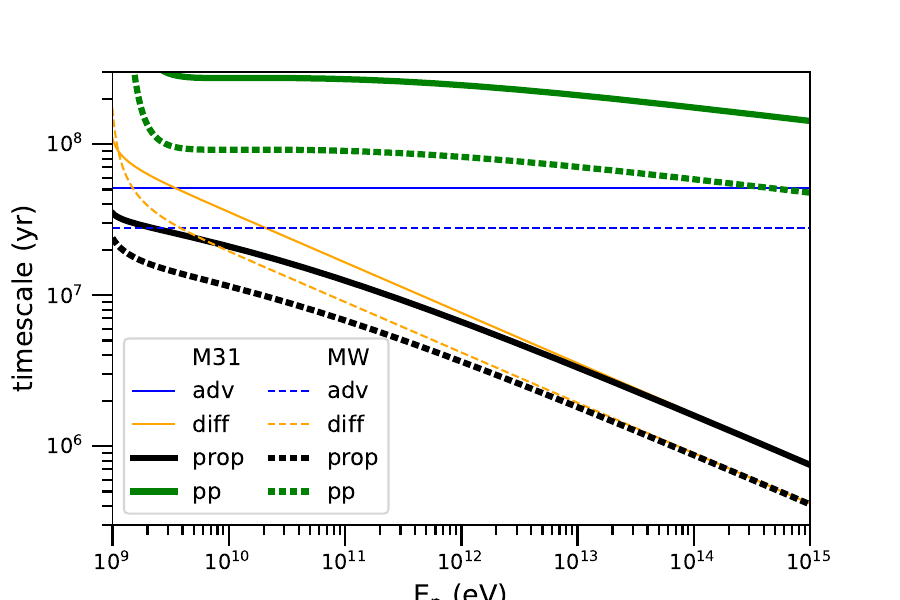}
\put(15,62){(a)}
\end{overpic}
\end{minipage}
\begin{minipage}{0.5\textwidth}
\centering
\begin{overpic}[width=\textwidth]{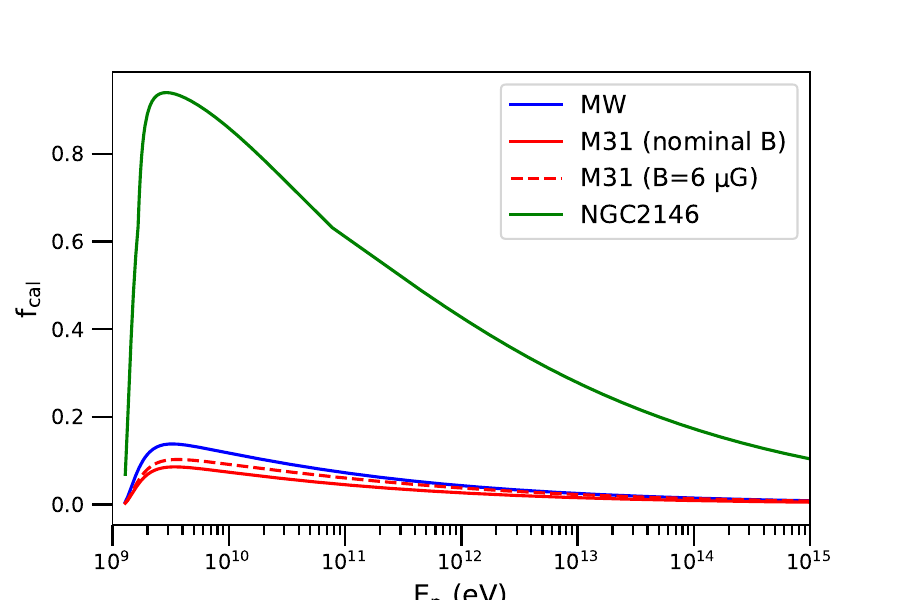}
\put(15,62){(b)}
\end{overpic}
\end{minipage}
\\
\\
\begin{minipage}{1\textwidth}
\centering
\begin{overpic}[width=0.8\textwidth]{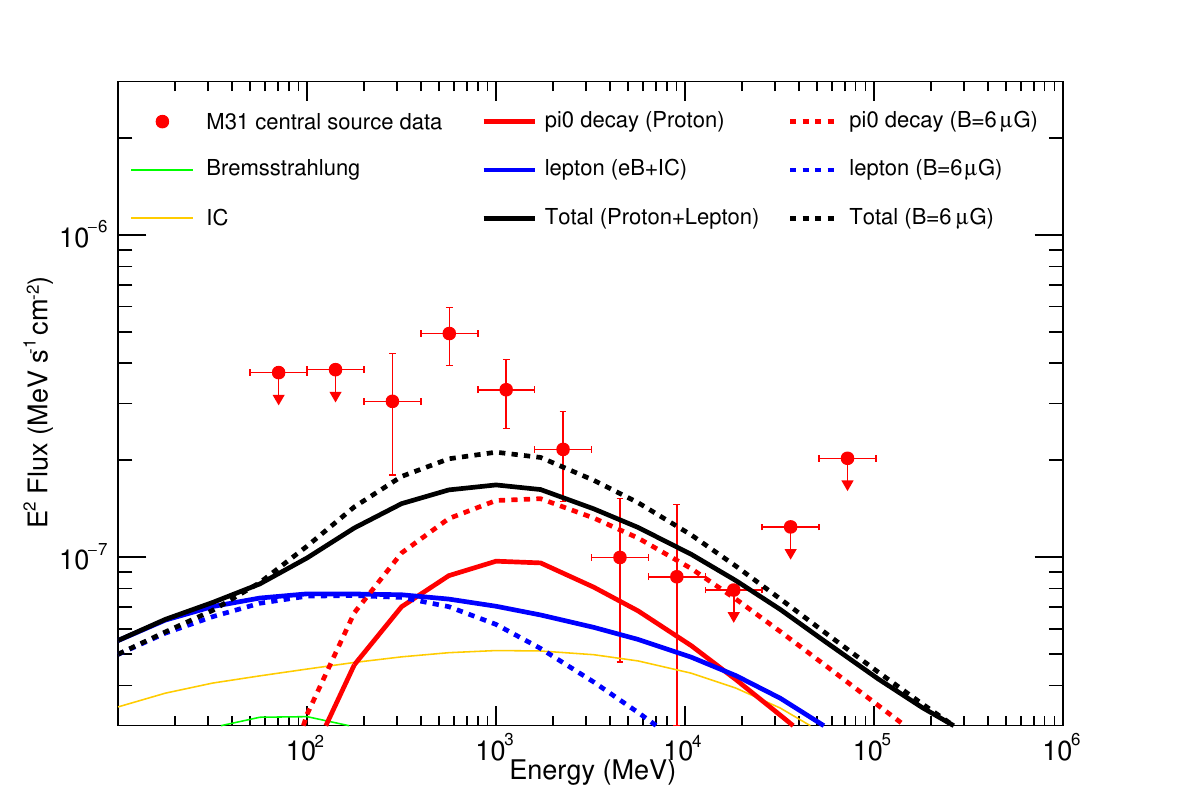}
\put(12,64){(c)}
\end{overpic}
\end{minipage} 
\end{tabular}
\caption{(a) Timescales of CR proton interaction and propagation of M31. 
For comparison, those of the Milky Way are overlaid.  The net propagation time $t_\mathrm{pr}$, 
defined as $(t_\mathrm{diff}^{-1} + t_\mathrm{adv}^{-1})^{-1}$, is also presented.
(b) Calorimetric efficiency of M31, Milky Way, and NGC~2146.
(c) SED of the M31 central source and the model prediction of interstellar $\gamma$ rays.
{Alt text: Three plots to show CR proton timescales, calorimetric efficiencies, and the SED. All include model curves for M31.}
}
\end{figure}

\clearpage 

\subsection{CR Production and Propagation of SFGs/SBGs}
In SFGs and SBGs, interstellar $\gamma$ rays contribute significantly to the total $\gamma$-ray emission. This contribution is substantial in SFGs and predominant in SBGs. Figure~14 presents the relationship between infrared luminosity ($L_\mathrm{IR}$) and $\gamma$-ray luminosity ($L_\gamma$) for the selected targets. Specifically, $\gamma$-ray luminosity above 100~MeV is plotted as a function of infrared luminosity in the 8--1000~$\mu$m band, as reported by \citet{Ajello2020}.
As discussed in Section~4.2, the contribution from unresolved sources is considerable
due to the limited spatial extent of M31's central source.
In contrast, most of the potential interstellar $\gamma$-ray emission,
which is expected to be symmetric about the galaxy's center, 
will be absorbed by the central source model below 1~GeV, where
angular resolution is moderate. 
Therefore, 
the observed $L_\gamma$ of M31's central source is treated
as an upper limit for interstellar $\gamma$ rays. 
As detailed in Section~3.4, the $\gamma$-ray emission from M33 is offset by ${\sim}0\fdg14$ from the galaxy's center and is associated with the star-forming region NGC~604.
Since NGC~604 is 3--6~Myr old (e.g., \cite{Relano2009}), which is considerably younger than the typical interaction time of CR protons in SFGs in the GeV range (>10~Myr; see Figure~13(a)), 
point source contributions are likely to be significant.
Consequently, the observed $L_\gamma$ is also considered an upper limit. For comparison, data points from \citet{Ajello2020} for the selected targets, as well as for the LMC,\footnote{
\citet{FermiLMC2} spatially resolved LMC emission into five components.
\citet{Ajello2020} evaluated the $\gamma$-ray emission of the large-scale disk component
(labeled E0) to study the interstellar $\gamma$-ray emission, and we adopted
their results. \citet{Persic2022} did a similar procedure for their multiwavelength modeling of the LMC.}
SMC, and Milky Way, are included.
A model curve for the CR calorimetric limit from \citet{FermiSFG} is also presented.

Our new analyses substantially reduce the $\gamma$-ray luminosities for NGC~2146 and Arp~220 
by about a factor of 2, aligning all SBGs with or below the calorimetric limit,
with observed $f_\mathrm{cal}$ values (ratio of the measured $\gamma$-ray luminosity to the calorimetric limit)
ranging from 0.33 to 1.20.
To verify this, we computed the theoretical $f_\mathrm{cal}(E_\mathrm{p})$ for NGC~2146 using the same procedure adopted for M31 and Milky Way (see Table~7 for model parameters).
The resulting efficiency exceeds 0.8 at a few GeV (Figure~13(b)), 
which is consistent with the observed value of 0.65. 
These findings demonstrate that interstellar $\gamma$-ray emission dominates all SBGs
in our sample.
The modest variations in $f_\mathrm{cal}$ among our sources
will enable accurate predictions of the starburst contribution
to both the extragalactic $\gamma$-ray background and the $\gamma$-ray emission of individual galaxies;
the latter allows distinguishing nuclear emission from the starburst component in Seyfert galaxies.
In addition, the SEDs obtained for M82 and NGC~253 will provide observation-based templates for future studies.

In contrast, all SFGs presented in Figure~14 exhibit low $f_\mathrm{cal}$ values at the 10\% level, 
except for the SMC, which has $f_\mathrm{cal}=0.28$.
As a low-mass dwarf galaxy, the SMC is expected to show a 
significant contribution from individual point sources.
Among the SFGs analyzed, the Milky Way displays unique $\gamma$-ray characteristics.
Although its $f_\mathrm{cal}$ is only {$\sim$}0.1, it remains dominated by interstellar $\gamma$-ray emission (e.g., \cite{Acero2016}),
due to its relatively high star-formation rate, which is
within a factor of 2 to 4 of those observed in the SBGs NGC~253 and M82.

Regarding the CR property of individual galaxies, our detailed SED fits (Section~4.1)
provide constraints on $K_\mathrm{ep}$. The inferred $K_\mathrm{ep}$ values of
M82 and NGC~253 ($1/25$--$1/30$ at 1~GeV and $1/120$--$1/150$ at 10~GeV) are 
comparable to those observed in the Milky Way (e.g., \cite{Gabici2022}).
The higher $K_\mathrm{ep}$ for M82, previously inferred by \citet{VERITAS_M82v2},
is no longer necessary given the revised SED below a few hundred MeV.
In scenarios invoking the diffusive shock acceleration at supernova remnants (SNRs),
$K_\mathrm{ep}$ is generally expected to be universal
(e.g., \cite{Merten2017}), and a value
about $1/100$ is frequently adopted to predict
$\gamma$-ray emission from SFGs and SBGs (e.g., \cite{Yoast-Hull2013}; \cite{Chen2025}).
Despite uncertainties arising from potential energy losses of CR electrons during acceleration or escape,
as well as contributions from CR accelerators other than SNRs,
the current results offer observational support for the canonical $K_\mathrm{ep}$ value.

Turning to the spectral indices, we obtained 2.4--2.5 
and ${\sim}$2.0 for $\alpha_\mathrm{p}$ and $\alpha_\mathrm{e}$, respectively (see Table~6). The latter gives the spectral index for CR electrons of about 3.0
above $E_\mathrm{bk2}$.
Consequently, both CR protons and electrons show harder spectral indices than those observed in the Milky Way, consistent with expectations for a system characterized by high CR calorimetric efficiency rather than significant CR escape.

Lastly, the CR proton energy densities ($U_\mathrm{CR}$) of M82 and NGC~253 are calculated from $W_\mathrm{p}$ and the 
starburst core size, yielding values of $\sim 190~\mathrm{eV\,cm^{-3}}$ and $\sim 470~\mathrm{eV\,cm^{-3}}$, respectively.
We remind that an independent analysis by \citet{Persic2025} gave similar $U_\mathrm{CR}$ for M82.
Our values are {$\sim$}8 and {$\sim$}2.5 times smaller than the magnetic-field energy densities ($U_\mathrm{mag}$). 
Although the magnetic field strength $B$ is subject to significant uncertainty,
reducing that of M82 by a factor of two (which would decrease
$U_\mathrm{mag}$ by a factor of four) prevents a satisfactory fit to the multiwavelength SED.
In this scenario, 
the inferred CR electron spectrum required to reproduce the radio data becomes too intense
and leads to an overprediction of $\gamma$-ray emission below a few hundred MeV.
Therefore, we conclude that the $U_\mathrm{CR}$ of SBGs,
at least for M82 and NGC~253, is well below energy equipartition, in contrast to the Milky Way. 
This could be due to the strong loss of CR protons (via interactions with interstellar gas)
and CR electrons (via interactions with ISRF and/or magnetic field) in dense environments,
as discussed by, e.g., \citet{Crocker2021}.
We note that the assumption of
energy equipartition between CRs and the magnetic field
is frequently used to estimate $B$ of SBGs from synchrotron emission (e.g., \cite{Adebahr2017,Persic2010}),
and these estimates may be subject to bias.

\begin{figure}[htbp]
\begin{tabular}{cc}
\begin{minipage}{1\textwidth}
\centering
\begin{overpic}[width=0.8\textwidth]{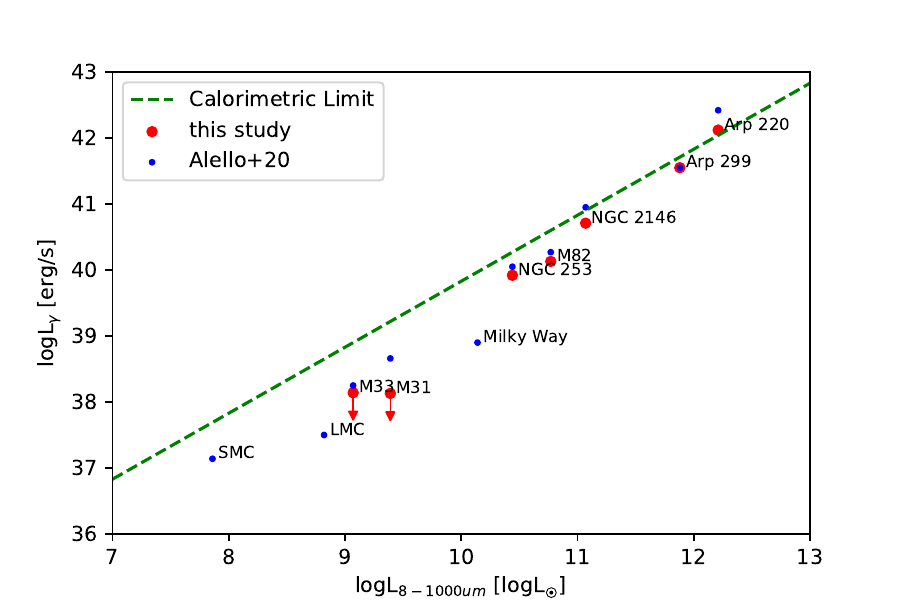}
\end{overpic}
\end{minipage} 
\end{tabular}
\caption{$\gamma$-ray luminosity vs. infrared luminosity
of our targets and the SMC, the LMC, and the Milky Way.
{Alt text: A plot to compare $\gamma$-ray lulminosity and infrared luminosity.}
}
\end{figure}

\clearpage

\section{Summary and Prospect}

We analyzed 15 years of Fermi-LAT data to characterize GeV and sub-GeV $\gamma$-ray emission from seven SFGs and SBGs, aiming to investigate
their CR properties and interstellar environments. The galaxies studied
include M82, NGC~253, M31, Arp~220, NGC~2146, Arp~299, and M33. While these galaxies were previously detected in GeV $\gamma$-rays, our improved data analysis, 
incorporating additional data and updated Galactic diffuse background and source models, enables a more detailed investigation of their $\gamma$-ray emission.
We found that the two bright SBGs, M82 and NGC~253, exhibit mild spectral curvature.
Our updated SEDs show a clear decrease in energy flux below a few hundred MeV, 
indicating a substantially smaller contribution from CR electrons than previously inferred. Through detailed multiwavelength SED fitting,
spanning radio to TeV $\gamma$-ray energies,
we determined that 
CR electron-to-proton ratios
are comparable to that of the Milky Way.
However, the CR energy densities in these two SBGs are significantly lower than
the energy equipartition with the magnetic field energy density, in contrast to the Milky Way.
For M31, the $\gamma$-ray emission consists of two point-like sources,
with the central source displaying a soft spectrum and a compact size of less than $0\fdg33$,
suggesting a considerable contribution from unresolved sources in addition to interstellar $\gamma$-ray emission.
Our revised analysis also yields lower $\gamma$-ray flux estimates for several SBGs.
The updated $\gamma$-ray luminosities of our SBG sample members are comparable 
to or below the calorimetric limit, with calorimetric efficiencies ranging from 0.33 to 1.20. In contrast, all our SFG sample members exhibit low efficiencies at the 10\% level,
except for the SMC, where point source contributions are likely significant. 
We further developed a theoretical model for interstellar $\gamma$-ray emission in SFGs and SBGs, 
which successfully reproduces the observed 
calorimetric efficiencies of NGC~2146 (SBG) and M31 (SFG).
Collectively, these results provide a robust framework for interpreting and predicting interstellar $\gamma$-ray emission from SFGs and SBGs.

\clearpage
\begin{ack}
The \textit{Fermi} LAT Collaboration acknowledges generous ongoing support
from a number of agencies and institutes that have supported both the
development and the operation of the LAT as well as scientific data analysis.
These include the National Aeronautics and Space Administration and the
Department of Energy in the United States, the Commissariat \`a l'Energie Atomique
and the Centre National de la Recherche Scientifique / Institut National de Physique
Nucl\'eaire et de Physique des Particules in France, the Agenzia Spaziale Italiana
and the Istituto Nazionale di Fisica Nucleare in Italy, the Ministry of Education,
Culture, Sports, Science and Technology (MEXT), High Energy Accelerator Research
Organization (KEK) and Japan Aerospace Exploration Agency (JAXA) in Japan, and
the K.~A.~Wallenberg Foundation, the Swedish Research Council and the
Swedish National Space Board in Sweden.
 
Additional support for science analysis during the operations phase is gratefully
acknowledged from the Istituto Nazionale di Astrofisica in Italy and the Centre
National d'\'Etudes Spatiales in France. This work performed in part under DOE
Contract DE-AC02-76SF00515.

\end{ack}

\section*{Funding}
Part of this work was supported by JSPS KAKENHI Grant Numbers 23K25882 and 23H04895 (T.M.), 26H00822 and 26H02063 (T.T.), and by JST SPRING Grant Number JPMJSP2108 (J.C.). \\ 

\section*{Data availability} 
The data underlying this article will be shared on reasonable request to the corresponding author.

\clearpage
\appendix 

\section{SEDs of Weak objects}
In Figure~15, we give SEDs of weak objects obtained using logP model.

\begin{figure}[htbp]
\begin{tabular}{cc}
\begin{minipage}{0.5\textwidth}
\centering
\begin{overpic}[width=\textwidth]{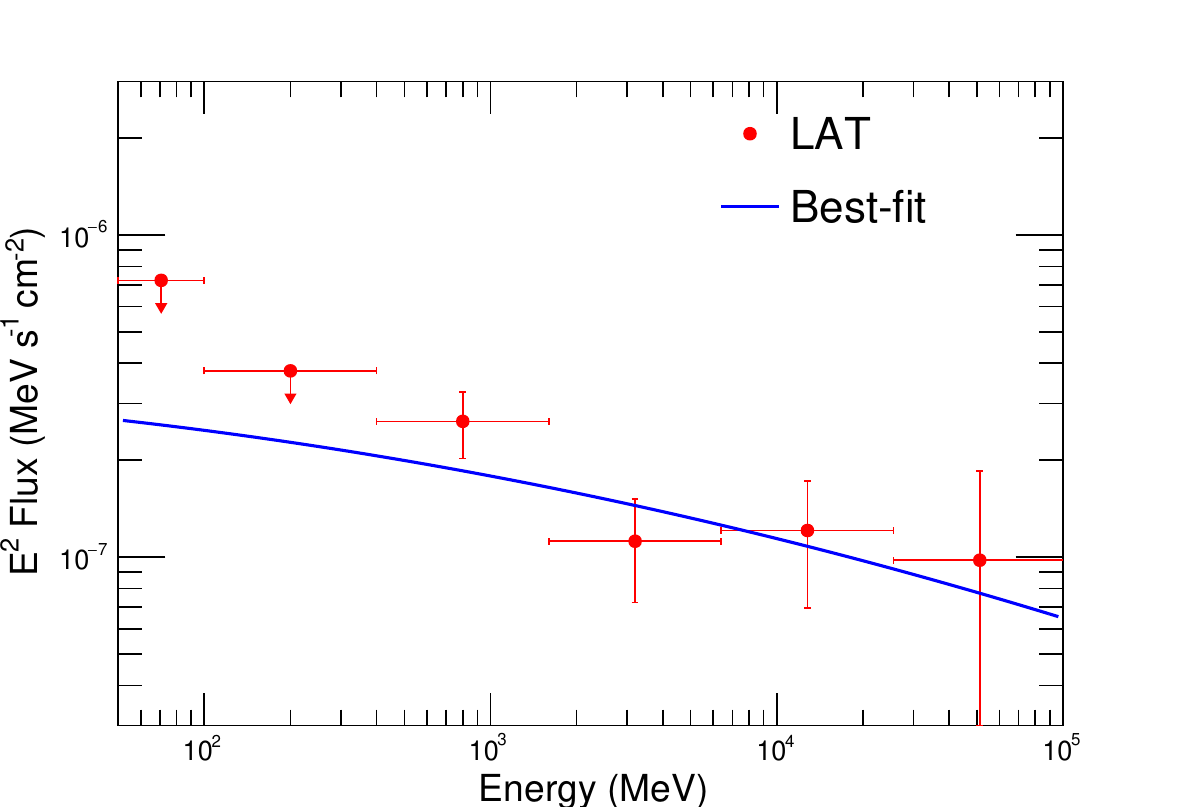}
\put(11,65){(a) Arp~220}
\end{overpic}
\end{minipage}
\begin{minipage}{0.5\textwidth}
\centering
\begin{overpic}[width=\textwidth]{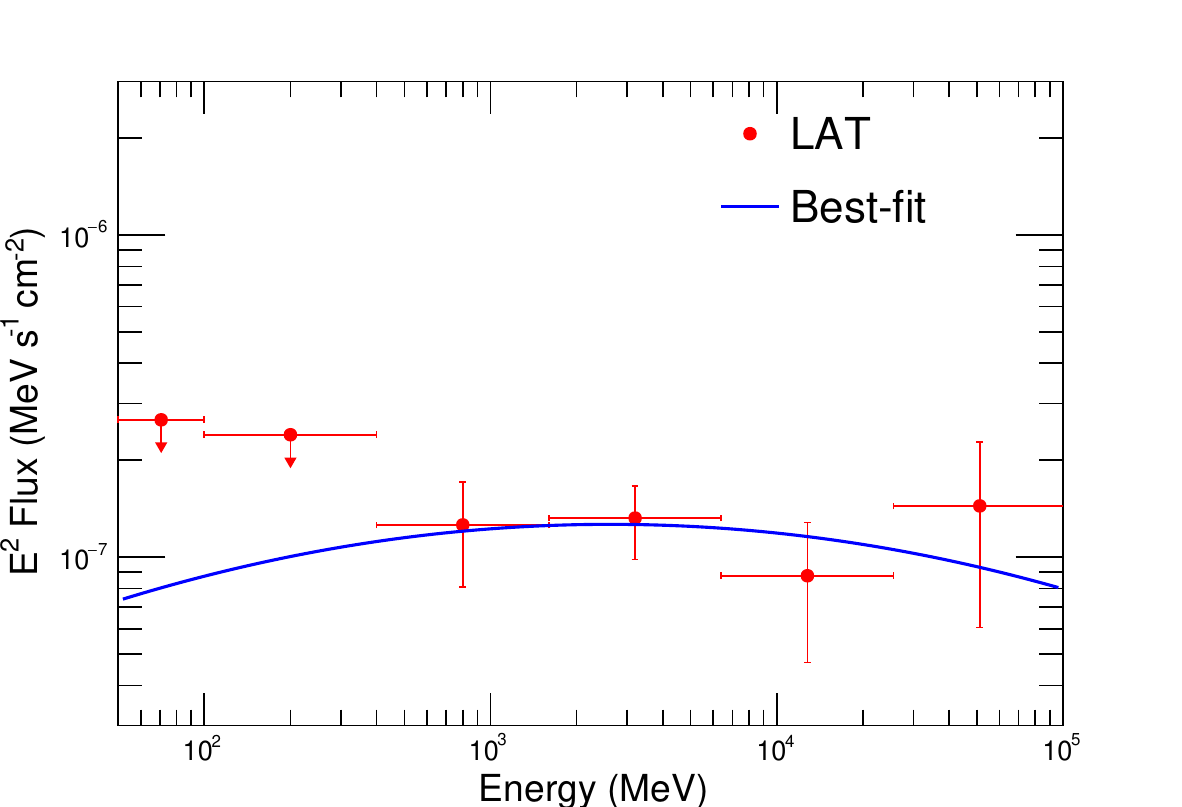}
\put(11,65){(b) NGC~2146}
\end{overpic}
\end{minipage} 
\\
\\
\begin{minipage}{0.5\textwidth}
\centering
\begin{overpic}[width=\textwidth]{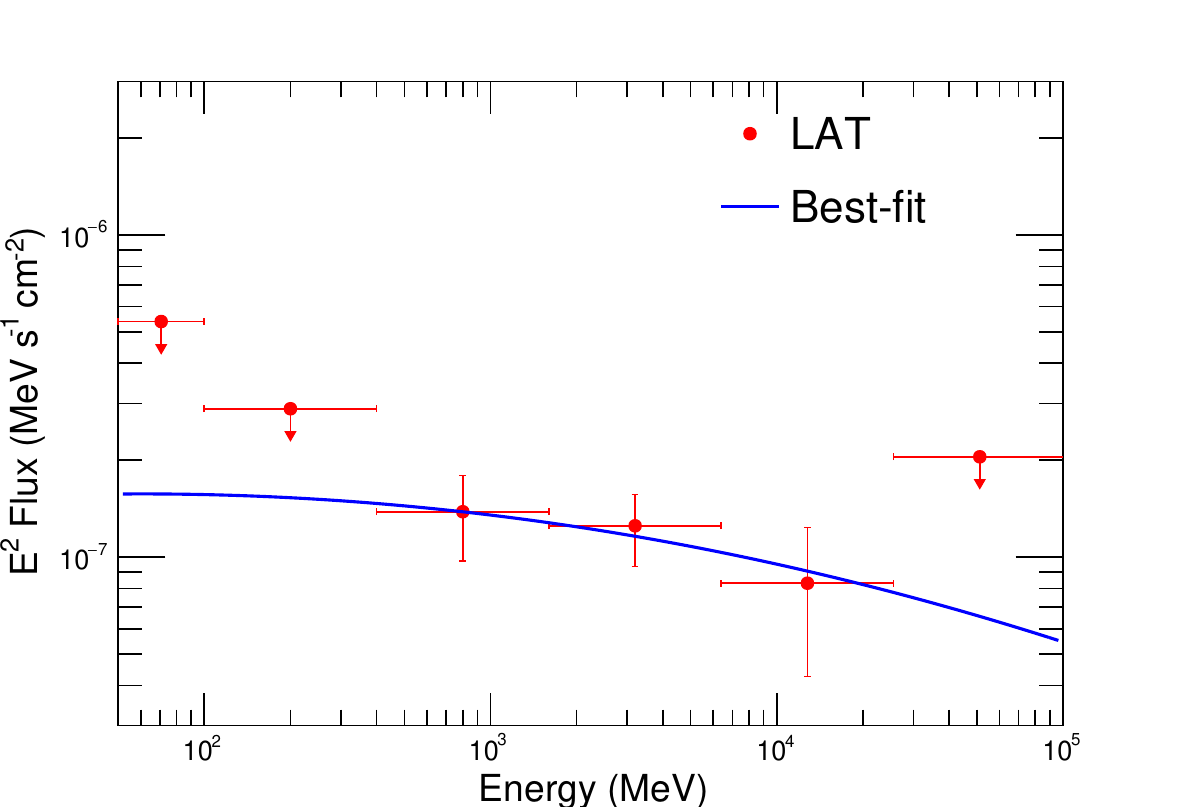}
\put(11,65){(c) Arp~299}
\end{overpic}
\end{minipage}
\begin{minipage}{0.5\textwidth}
\centering
\begin{overpic}[width=\textwidth]{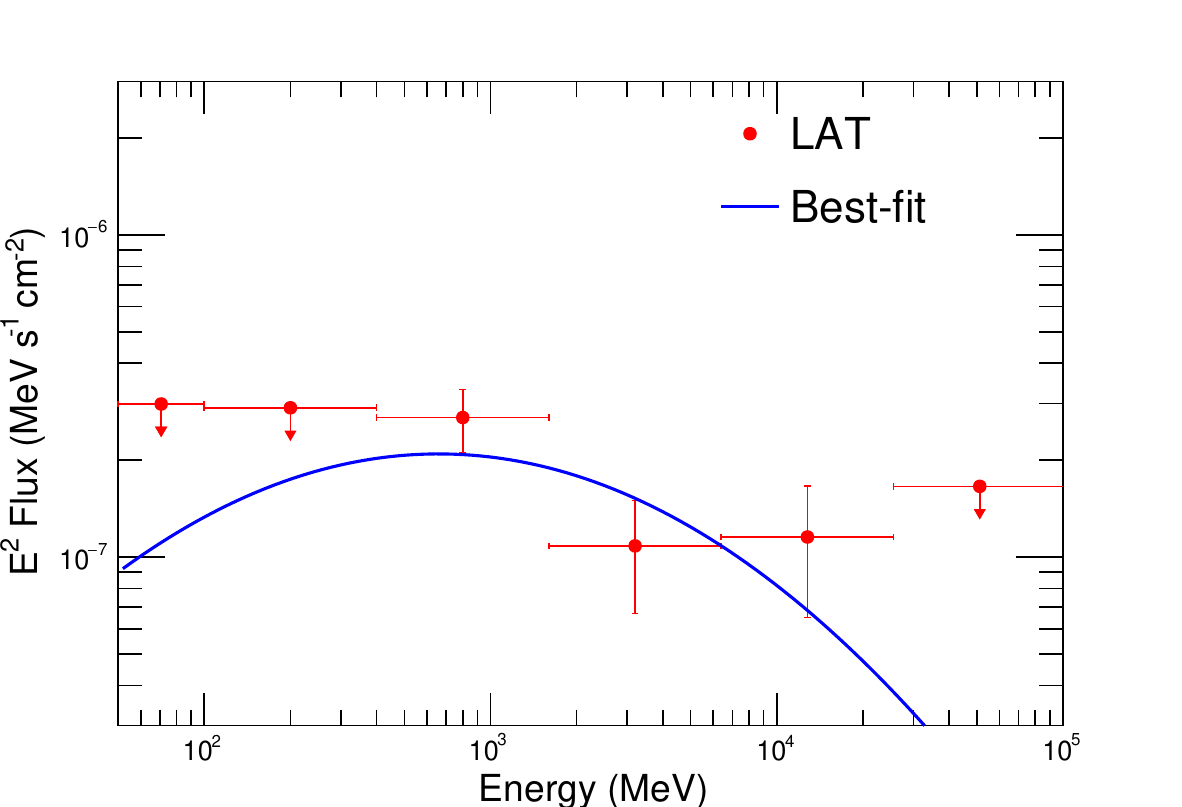}
\put(11,65){(d) M33}
\end{overpic}
\end{minipage}
\end{tabular}
\\
\caption{SED and a best-fit spectral model of (a) Arp~220, (b) NGC~2146,
(c) Arp~299, and (d) M33, obtained by the logP model.
{Alt text: Four SED plots for other weak objects.}
}
\end{figure}
\clearpage

\section{Theoretical Modeling of Gamma-ray SED}
In order to model the SED of the M31's central source (the flux is approximately half that of M82 and NGC~253), 
we employ the model of $\gamma$-ray emission from SFGs and SBGs developed by \citet{Chen2025} (hereafter C25) based on \citet{Sudoh2018} and \citet{Shimono2021}.
This model calculates $\gamma$-ray emission from CR protons and electrons based on several key galactic parameters: star formation rate $\psi$, gas mass $M_{\mathrm{gas}}$, stellar mass $M_{\mathrm{star}}$, effective radius $R_{\mathrm{eff}}$, and luminosities in optical and infrared bands ($L_{\mathrm{opt}}$ and $L_{\mathrm{IR}}$.), assuming a one-zone disk-like geometry. 
However, the model used here
has undergone several significant improvements compared to C25, which are described below.

In the models up to C25, it was assumed that CR particles are confined within the gas scale height $H_{\rm gas}$ from the Galactic plane; however, in more realistic CR transport models (e.g., GALPROP; \cite{Galprop1,Galprop2}),
the CR scale height $H_{\rm CR}$ is considerably larger than $H_{\rm gas}$. 
This is particularly important to CR electrons interacting with the radiation field extending from the gas disk. 
Therefore, in the new model, the scale height of the cosmic ray distribution is determined based on physical considerations as follows.
Let $H_{\rm halo}$ denote the scale height of the geometric boundary at which cosmic rays escape beyond the galaxy. 
In the Milky Way, which serves as the model reference, 1 kpc is adopted for $H_{\rm halo}$ from a typical scale height for hot gas, and $H_{\rm halo}$ in other galaxies is determined on the assumption that the ratio of $H_{\rm halo}/R_{\rm eff} = 1/6$ is constant across all galaxies.

For a given energy of CR protons, we define $H_{\rm CR,p}$ as the scale height at which CR protons propagate without losing a significant amount of their energy.
The time $t_{\rm pr,p}$ to propagate the distance $H_{\rm CR,p}$ is determined by the shorter of the two time scales associated with diffusion ($t_\mathrm{diff}$) 
and advection ($t_\mathrm{adv}$), i.e., $t_{\rm pr,p} = \left(t_{\mathrm{diff}}^{-1}+t_{\mathrm{adv}}^{-1}\right)^{-1}$, where two time scales are calculated by the formulation of C25. 
Here, it is assumed that the same formulation for diffusion in C25 can be applied to the height $H_{\rm CR,p}$, beyond $H_{\rm gas}$. This is not unreasonable, provided that energy is roughly equally distributed between CRs and magnetic fields by strong coupling.
If the spread of CR protons is limited by energy loss due to the pp reaction, the scale height will be $\tilde H_{\rm CR,p}$, which is determined by $t_{\rm pr,p}(H_{\rm CR,p}) = t_{\rm pp}(H_{\rm CR,p})$, where $t_{\rm pp} = (H_{\rm CR,p}/H_{\rm gas}) (n_{\rm gas} \sigma_{\rm pp} v_{\rm p})^{-1}$, and $v_{\rm p}$ is the velocity of protons.\footnote{In C25, formulas of $t_{\rm pp}$, the Larmor radius, and the diffusion coefficient (eq. 4) were based on the approximation that the proton's velocity is nearly the speed of light. Regarding these, corrections were made such as replacing $c$ with $v_{\rm p}$.} 
The $(H_{\rm CR,p}/H_{\rm gas})$ factor takes into account the proportion of time that protons spend in the gas disk. The final $H_{\rm CR,p}$ is determined by the following equation:
\begin{eqnarray}
H_{\rm CR,p} = \left\{
\begin{array}{cc}
   H_{\rm gas}  &  ({\rm if} \ \tilde H_{\rm CR,p} < H_{\rm gas} ) \\
   \tilde H_{\rm CR,p}  &  ({\rm if} \ H_{\rm gas} \le \tilde H_{\rm CR,p} \le H_{\rm halo})\\
   H_{\rm halo}  &  ({\rm if} \ H_{\rm halo} < \tilde H_{\rm CR,p})
\end{array}  \right. \ .
\label{eq:H_CRp}
\end{eqnarray}

Due to significant uncertainties regarding the diffusion coefficients of CR protons and electrons, we will determine it so as to reproduce the mean path length (grammage)  $\lambda = M_{\rm gas} /(2\pi R_{\rm eff}^2 H_{\rm CR,p}) \, v_{\rm p} \, t_{\rm pr,p}= 10 \,\mathrm{g/cm^2}$ for CR protons at $E$ = 1 GeV, which is a typical value for the Milky Way. 
We find that we can reproduce this $\lambda$ value by applying a correction factor of 2.9 to the formula of the C25 diffusion coefficient; therefore, we apply this correction factor to all galaxies and cosmic ray energies.
In this case, 1 GeV CR protons in the Milky Way are non-calorimetric ($H_{\rm CR,p} = H_{\rm halo}$), and the escape time from the CR halo is $1.4\times10^7$ years (see Figure~13(a)).

It should be noted that the galaxy becomes non-calorimetric to CR protons only in the final case of eq. (\ref{eq:H_CRp}), i.e., 
$H_{\rm halo} < \tilde H_{\rm CR,p}$.
Then the gamma-ray luminosity is calculated using the calorimetric efficiency $f_{\mathrm{cal}} = 1 - \exp\left[- t_{\rm pr,p}/ t_{\mathrm{pp}}\right]$.
Conversely, when $\tilde H_{\rm CR,p} \le H_{\rm halo}$, CR protons experience substantial energy loss before reaching $H_\mathrm{halo}$, resulting in a calorimetry of the galaxy. In this case, since $t_{\rm pr,p} = t_{\mathrm{pp}}$ always, $f_\mathrm{cal}$ is calculated as $f_{\mathrm{cal}} = 1 - \exp\left[- H_{\rm halo}/ H_{\mathrm{CR,p}}\right]$, such that it connects continuously to the non-calorimetric $f_{\rm cal}$ when $\tilde H_{\rm CR,p} = H_{\rm halo}$, 
and $f_\mathrm{cal}$ asymptotically approaches 1
when $\tilde H_{\rm CR,p} \ll H_{\rm halo}$.

MW and M31 are non-calorimetric, but NGC 2146 is an example of a calorimetric galaxy, as illustrated in Figure 16. 
Here, various time scales for this galaxy are shown in Figure~16(a), in comparison to those of MW and M31 in Figure~13(a).
Figure~16 (b) summarizes $H_\mathrm{CR, p}$ of these three galaxies.
From a few GeV to a few tens GeV, $H_\mathrm{CR, p}$ of NGC 2146 is lower than $H_\mathrm{halo}$, meaning that this galaxy is calorimetric in this energy range.

The scale height of CR electrons, $H_{\rm CR,e}$, is determined in a manner similar to that for protons.
Propagation by diffusion or advection is the same as that of protons.
The energy loss timescale for electrons is the cooling time $t_{\rm cl, e}$ due to ionization, bremsstrahlung, and synchrotron and IC radiation.
Here, regarding the energy loss rates due to ionization and bremsstrahlung, the factor $H_{\rm gas}/H_{\rm CR,e}$ is applied to that of C25 to account for the residence time within the gas disk.
On the other hand, the energy losses due to synchrotron and IC radiation are taken from C25 without modification.
This is a reasonable assumption, given that the radiation field from the disk persists without significant attenuation up to the halo scale height, $H_{\rm halo} = (1/6) R_{\rm eff}$, and that cosmic rays and magnetic fields are strongly coupled.
See \cite{kalberla05} and \cite{Fukui2017} for details.

\begin{figure}[htbp]
\begin{tabular}{cc}
\begin{minipage}{0.5\textwidth}
\centering
\begin{overpic}[width=\textwidth]{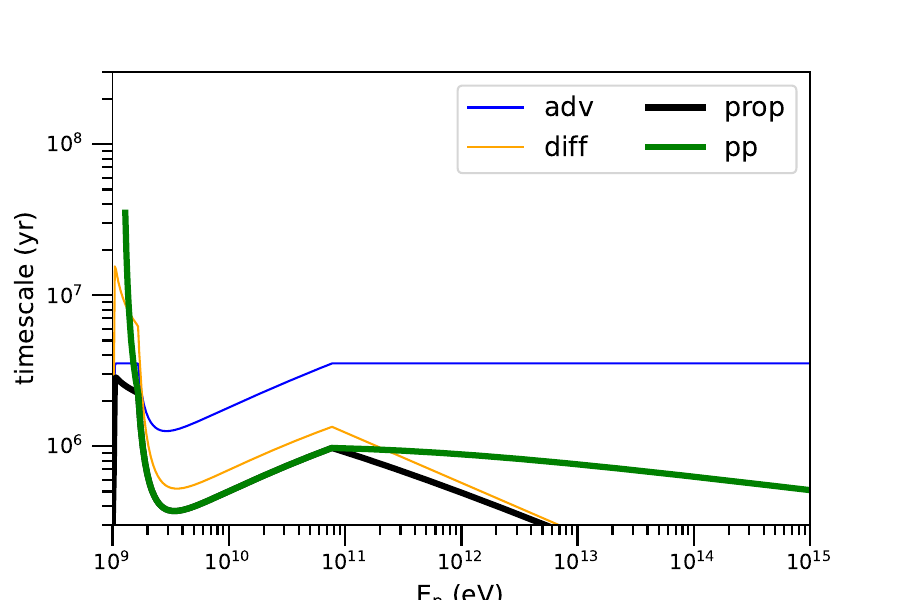}
\put(15,62){(a)}
\end{overpic}
\end{minipage}
\begin{minipage}{0.5\textwidth}
\centering
\begin{overpic}[width=\textwidth]{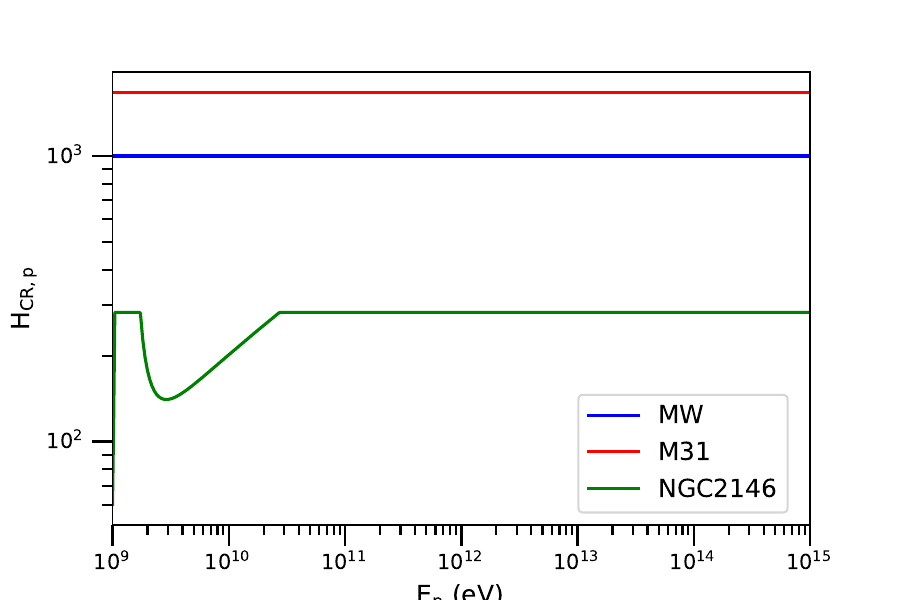}
\put(15,62){(b)}
\end{overpic}
\end{minipage}
\end{tabular}
\caption{(a) Timescales of CR proton interaction and propagation of NGC~2146. 
(b) $H_\mathrm{CR, p}$ of three galaxies. They are equal to
$H_\mathrm{halo}$ except $E_\mathrm{p}$ from a few GeV to a few tens GeV for NGC~2146.
{Alt text: Two plots to show CR proton timescales and calorimetric efficiencies. All include model curves for NGC~2146.}
}
\end{figure}
\clearpage








\bibliography{references}{}
\bibliographystyle{aasjournalv7}

\end{document}